\documentclass[%
 aip,
 amsmath,amssymb,
]{revtex4-1}

\usepackage{graphicx}
\usepackage{subfigure}
\usepackage{ulem}
\usepackage{dcolumn}
\usepackage{bm}

\usepackage[utf8]{inputenc}
\usepackage[T1]{fontenc}
\usepackage{mathptmx}
\usepackage{xcolor}
\usepackage{soul}

\begin{document}


\title[]{Onset of transient shear banding in viscoelastic shear start-up flows: Implications from linearized dynamics}
\author{Shweta Sharma}
\author{V.~Shankar}%
 \email{vshankar@iitk.ac.in}
 \author{Yogesh M. Joshi}
 \email{joshi@iitk.ac.in}
\affiliation{Department of Chemical Engineering, Indian Institute of Technology Kanpur, Kanpur 208016, India
}%
\begin{abstract}
 
We analyze transient dynamics during shear start-up in viscoelastic flows between two parallel plates, with a specific focus on the signatures for the onset of transient shear banding using the Johnson-Segalman, non-stretching Rolie-Poly and Giesekus models. We explore the dynamics of shear start-up in monotonic regions of the constitutive curves using two different methodologies: (i) the oft-used `frozen-time' linear stability (eigenvalue) analysis, wherein we examine whether infinitesimal perturbations imposed on instantaneous stress components (treated as quasi steady states) exhibit exponential growth, and (ii) the more mathematically rigorous fundamental-matrix approach that characterizes the transient growth via a numerical solution of the time-dependent linearized governing equations, wherein the linearized perturbations co-evolve with the start-up shear flow. Our results reinforce the hitherto understated point that there is no universal connection between the overshoot and subsequent decay of shear stress in the base state and the unstable eigenvalues obtained from the frozen-time stability analysis. It may therefore be difficult to subsume the occurrence of transient shear banding during shear start-up within the ambit of a single model-independent criterion. Our work also suggests that the strong transients during shear start-up seen in earlier work could be sensitive to the small solvent viscosity values considered in the absence of otherwise negligible terms such as fluid inertia.

\end{abstract}
\maketitle
\section{Introduction}
Industrial processing and rheological characterization of soft materials are often susceptible to flow instabilities of various kinds, such as wall slip [\onlinecite{denn2001extrusion,joshi2000slipping}], shark-skin [\onlinecite{petrie1976instabilities,howells1962flow,malkin2006flow}], hydrodynamic  [\onlinecite{Shaqfeh_1996,larson1992}], and shear banding instabilities [\onlinecite{olmsted2008perspectives,divoux2016shear,germann2019shear}]. Shear banding is the phenomenon wherein an initially linear velocity profile (between two parallel plates) with a homogeneous shear rate becomes unstable resulting in the formation of coexisting bands, each having a different shear rate. The two bands are arranged along the wall-normal direction, with the interface between the two bands oriented along the flow-vorticity plane. 
 In this study, we focus on the bulk shear banding instability in a shear start-up flow of viscoelastic polymeric or wormlike micellar solutions. 
It is known that steady state shear banding is a constitutive instability which arises because of the inherent non-monotonic nature of the constitutive curve of a material [\onlinecite{yerushalmi1970stability}]. A sufficient condition for the onset of shear banding instability is the presence of a negative slope in the stress-strain rate constitutive curve. This criterion is independent of the details of the constitutive relation, and hence could be regarded as `fluid-universal'. In the non-monotonic regions of the constitutive curve, an initially homogeneous, steady velocity profile  is linearly unstable, and the flow will evolve into an eventual banded steady state.
Non-monotonic constitutive curves are exhibited by various constitutive models such as the Johnson-Segalman (abbreviated `JS') model [\onlinecite{johnson1977model}], Doi and Edwards model [\onlinecite{doi1988theory}], Cates model [\onlinecite{cates1987reptation}], the `VCM' (Vasquez–
Cook–McKinley) model [\onlinecite{vasquez2007network}], Giesekus model [\onlinecite{giesekus1982simple}], and Rolie-Poly model [\onlinecite{likhtman2003simple}], etc. Recently, Ianniruberto and Marucci [\onlinecite{ianniruberto2017shear}] used an integral version of the Doi-Edwards constitutive equation
and demonstrated the linearly unstable nature of non-monotonic region of the constitutive curve.

Interestingly, shear banding has been reported to occur not only at steady state, but also during the transient evolution, the latter being labelled `transient shear banding'. 
While the term transient shear banding has been employed frequently in the literature, to our knowledge, there has not been an endeavor to precisely define the same. Here, we generalize the notion of steady-state shear banding to the transient case. Transient shear banding may be said to be present when two or more bands with distinctly different shear rates get formed transiently with a transition zone being much smaller compared to the width of the bands. However, this definition does not account for whether the steady state is banded or not. If the eventual steady state is a banded one, then one might expect the velocity distributions to develop a banding-like inhomogeneity \textit{en route} to this banded state. While this might be technically called transient banding, it is caused only because of evolution towards the eventual banded steady state. In such instances, transient banding may be said to occur only \textit{if} the intermediate flow states exhibit substantially pronounced banding than the eventual banded steady state [\onlinecite{moorcroft14}]. Thus, a more definitive way to ascertain transient banding in an unambiguous manner would be in instances where the eventual state is not banded (i.e., homogeneous), and yet, during the transient evolution, there is a formation of two (or more) distinct bands with different shear rates with a sharp transition zone between the bands. 

An important question in the field of transient shear banding during shear start-up is whether there exists a universal and simple criterion, along the lines of the aforementioned negative-slope criterion for steady-state banding [\onlinecite{yerushalmi1970stability}], which will enable one to predict and explain the onset of transient shear banding using relatively simple rheological signatures. To this end, 
Moorcroft and Fielding [\onlinecite{moorcroft2013criteria,moorcroft14}]  proposed fluid-universal criteria for the onset of  transient shear banding under different flow protocols. Fielding [\onlinecite{fielding2016triggers}] noted that the criterion for transient banding for the step stress protocol could apply universally to all materials. However, as acknowledged by Fielding [\onlinecite{fielding2016triggers}], the criterion for shear start-up was derived only for models with two dynamic viscoelastic variables. In addition to this obvious restriction, there are other subtler assumptions that are involved in the derivation of the criterion for transient banding during shear start-up. These assumptions need to be understood clearly in order to ascertain the domains of applicability of the criterion. We discuss these aspects in detail below in Sec.~\ref{section_background}.

The goal of this work is  to study the (possible) onset of transient shear banding at length using three different viscoelastic constitutive models, and using two very different approaches within the realm of linearized dynamics. In the first approach, we carry out a classical linear stability analysis at each instant of time (henceforth referred to as the `frozen-time' analysis) and look for unstable eigenvalues, which would then be interpreted as signatures of a transient instability. In the second (and more rigorous) approach, we use the so-called fundamental matrix method wherein the perturbations co-evolve with the base-state, and the growth coefficient is calculated without making the frozen-time assumption. This method is analogous to what is traditionally referred to as `non-modal' stability analysis in the field of hydrodynamic stability theory [\onlinecite{schmid2007nonmodal}].
In Sec.~\ref{section_background}, we first assess the key issues involved in the formulation of a fluid-universal criterion for transient shear banding during shear start-up.  
By way of setting the context for the present study, we first provide a brief background of the relevant earlier work below. We then lay out the specific objectives of the present work at the end of Sec.~\ref{section_background}.

\section{Background and Objectives}\label{section_background}


One of the first studies that examined the stability of a time-evolving base-state, in the context of shear banding, was by Fielding and Olmsted [\onlinecite{fielding2003kinetics}] who carried out a linear stability analysis using the diffusive Johnson-Segalman model coupled with a micellar concentration field. The authors considered non-monotonic constitutive curves,  and explored both the steady state and the dynamical pathway to the steady state.  In their analysis, the authors found the possibility of a Hopf bifurcation, wherein two complex conjugate eigenvalues simultaneously become unstable, for their `Type II' system which is more susceptible to Cahn-Hilliard instability. 
Transient shear banding in shear start-up flow with shear rates in both monotonic and
nonmonotonic regions of the constitutive curve was also reported by Adams, Fielding and Olmsted [\onlinecite{adams2009nonmonotonic,adams2011transient}] using numerical computations of the the Rolie-Poly model.

In subsequent efforts by Moorcroft and Fielding [\onlinecite{moorcroft2013criteria,moorcroft14}], criteria were proposed for 
the onset of transient shear banding under different flow protocols. For shear start-up flow,  the criterion, in its simplest form, states that if there is an overshoot and subsequent decay of the shear stress as a function of strain, then an `elastic instability' will be triggered, which was proposed to presage  transient shear banding. A more accurate version of the criterion predicts the onset of banding at times a little before the occurrence of the overshoot.
Moorcroft and Fielding [\onlinecite{moorcroft14}] used numerical simulations and showed that the Rolie-Poly model exhibited transient shear banding. However, they noted the absence of the same in the shear start-up dynamics of the Giesekus model. This was attributed to the lack of convergence of the shear start-up curve to a limiting function of strain at high shear rates in the Giesekus model.  

In a follow-up pr\'{e}cis,
Fielding [\onlinecite{fielding2016triggers}] noted that there is good evidence in literature to indicate a good correlation between transient shear banding and stress overshoot. In the later part of the same article, Fielding nevertheless sounded a note of caution on the generality of the transient shear banding criterion [\onlinecite{moorcroft14}]: \textit{``Our most important caveat concerns the generality of the stress overshoot criterion for banding in shear startup. Although indeed derived in a constitutive model written in a generalized form, to make progress analytically, this allowed for only two dynamical viscoelastic variables.''}  The author also emphasized the need for further assessment of the transient shear banding criterion (during shear start-up) using more constitutive models and experiments [\onlinecite{fielding2016triggers}].In addition to the limitations pointed out by Fielding in the said pr\'{e}cis, 
during the course of the present work, we found that some of the other assumptions made in the analytical derivation of the criteria in Refs.~[\onlinecite{moorcroft14,moorcroft2013criteria}] related to transient banding in shear start-up might also have been restrictive. These assumptions raise further questions on the generality and fluid-universal nature of the criterion.
 Motivated by the above discussion of the literature, we identify and discuss below the reasons for the restrictive nature of the criterion for transient shear banding during start-up shear flow. The following discussion, therefore, forms a background to the earlier work  and also concurrently provides the motivation for the work carried out in the present manuscript.


First, in the steps preceding the derivation of the criteria for transient banding during shear start-up, the authors arrive at a linearized system of ordinary differential equations describing the evolution of the perturbations (Eq.~47 of Ref.~[\onlinecite{moorcroft14}]), of the form $\frac{\partial\mathbf{x}}{\partial t} = \mathbf{A} \cdot \mathbf{x}$, where $\mathbf{x}$ is the `state' vector comprising of the perturbation viscoelastic stress components, while $\mathbf{A}$ is the linearized matrix governing the evolution of the perturbations at a given time instant (crucially, this assumption is also restrictive, as discussed below, but we ignore this for the moment). Second, according to the authors, if at least one of the eigenvalues of $\mathbf{A}$ has a positive real part, then the perturbations will exhibit an exponential growth. The authors argue that if the instantaneous dynamical variables are stable (corresponding to all the eigenvalues of matrix $\mathbf{A}$ being negative, assuming them to be real eigenvalues), then the determinant $(-1)^D \det {\bf A} > 0$, while 
 $(-1)^D \det {\bf A} < 0$ if the system is unstable, 
 where $D$ is the order of ${\bf A}$. To arrive at this criterion, the authors first assume the existence of \textit{only one} positive \textit{real}, unstable eigenvalue, and explicitly ignore the possibility of a Hopf bifurcation, wherein a pair of complex conjugate eigenvalues become unstable as a parameter is varied [\onlinecite{Drazinreid}]. We show later in this work that this assumption breaks down in the JS model, where we demonstrate Hopf bifurcation in the linear stability analysis.
However, following earlier efforts [\onlinecite{moorcroft14}], we discount the possibility of Hopf bifurcation for the moment.  The matrix {\bf A} can be diagonalized by transforming it into its eigenbasis, with $\Lambda$ being a purely diagonal matrix with eigenvalues of $\bf{A}$ = $(\lambda_1 \cdots \lambda_D)$ as its diagonal entries.  The prefactor $(-1)^D$, mentioned above, accounts for the fact that $\det{\bf A} = \Pi_{i = 1}^{D} \lambda_i$, and if $D$ is odd (even), then $\Pi_{i = 1}^{D} \lambda_i$ would be positive (negative) when only one of the $\lambda$'s is unstable.  Having made the assumption of a single (real) eigenvalue becoming unstable, 
the expression that emerges from their analysis (Eq.~51 of Ref. [\onlinecite{moorcroft14}]) is rather elaborate.  To simplify this expression further, the authors assume the limit of very fast flow, and further neglect the derivatives with respect to strain rate in the above-mentioned expression, the reason being that the response of the system would be largely elastic and hence independent of strain rate. Finally, the authors restrict their analysis to $D = 2$, viz., constitutive models with only two stress components as also pointed out by Refs. [\onlinecite{fielding2016triggers,ianniruberto2017shear}]. It is after making these assumptions that the authors arrive at the criterion of negative slope in the stress-strain curve during shear start-up.

Even if we were to set aside the possible shortcomings of the aforementioned mathematical simplifications used to arrive at the final criterion, it is important to  make a clear distinction between whether the purported criterion for transient banding during start-up shear flow is a \textit{necessary} or a \textit{sufficient} condition.  Since the criterion of Moorcroft and Fielding [\onlinecite{moorcroft14}] were obtained by assuming the existence of positive (unstable) eigenvalues first, it can at best serve  only as \textit{necessary, but not sufficient} condition. In other words, constitutive models that exhibit exponential growth during transient evolution, with only one eigenvalue crossing over to become unstable,  will satisfy the criteria. However, \textit{not} all constitutive models that satisfy the criteria will exhibit exponential growth and thence transient banding. We demonstrate this using linearized dynamics of shear start-up flow in this study explicitly by considering three widely used constitutive models, viz., the Johnson-Segalman, non-stretching Rolie-Poly and Giesekus model, which exhibit a stress maximum and decay upon start-up shear, but do not always exhibit signatures of transient banding.

Interestingly, an analogy may be drawn from classical hydrodynamic stability, wherein the celebrated Rayleigh's inflection point theorem states that the presence of an inflection point in the velocity profile is a \textit{necessary} condition for instability of rectilinear shear flows in the inviscid limit [\onlinecite{Drazinreid}]. However, since the Rayleigh theorem is a necessary condition, it cannot be used to show that any given flow that satisfies the condition is unstable, but can only be used to show that the flows that do not satisfy the inflection point theorem are stable. For example, plane Couette and Poiseuille flows are stable in the inviscid limit because their velocity profiles do not posses a point of inflection.  In a similar vein, it can be argued that the criteria of Moorcroft and Fielding [\onlinecite{moorcroft14}], after acknowledging the above-mentioned limitations, may perhaps be used to \textit{rule out} transient banding during shear start-up in constitutive models that do not satisfy them, but not as a prescription to identify the presence of transient banding in constitutive models that do satisfy the criteria. However, even here,  as we discuss below, we must caution that there is the possibility of a transient growth of perturbations even when all the eigenvalues are stable.

There is a third, and perhaps more crucial aspect to the derivation of the criteria. The linearized evolution equations $\frac{\partial\mathbf{x}}{\partial t} = \mathbf{A} \cdot \mathbf{x}$ are amenable to a classical `normal mode'  temporal eigenvalue analysis with exponential growth or decay of perturbations only when the matrix $\mathbf{A}$ is independent of time. However, in the transient evolution of the stresses during start-up shear, the matrix $\mathbf{A}$ is necessarily a function of time. Thus, it is necessary to fix a time $t$ (the so-called `frozen-time' analysis) and carry out a normal-mode stability analysis in the usual sense [\onlinecite{von1974linear,davis1976stability}]. This assumption of freezing the time also places restrictions on the validity of the stability analysis, since the rate of growth or decay of perturbations must be very large compared to the rate of evolution of the base-state in order for the results to be self-consistent. 
While it is possible to provide specific initial conditions to the linearized governing equations and compute their evolution, there is a rather elegant method called the `fundamental matrix approach' [\onlinecite{alam1997influence,schmid1994optimal,schmid1994transient,strang1997linear}] which, using the principles of linear algebra, directly provides the growth (at each time) maximised over all possible initial conditions. 
It is pertinent here to note that, unlike the frozen-time approximation, the fundamental matrix analysis does not presuppose an exponential growth or decay. This method is analogous to the method of `non-modal stability' [\onlinecite{schmid2012}] used in the field of hydrodynamic stability. Indeed, it is well appreciated in hydrodynamic stability that perturbations to even steady flows can exhibit transient, non-exponential, growth before eventually decaying exponentially, even if the steady flow is stable as per modal (eigenvalue) analysis [\onlinecite{jovanovic2010transient}].  Whether this transient growth will lead to an instability of laminar flow, however, cannot be answered within the confines of a linearized analysis. Thus, the transient growth, when substantial, is usually taken as a plausible signature that nonlinear mechanisms can become dominant in making the flow unstable. In this work, we adopt the fundamental matrix approach to study the stability of time-evolving base states. We show that the results from the fundamental matrix approach are not always consistent with the predictions of the frozen-time eigenvalue analysis, further underscoring the restrictive nature of the criteria derived in literature [\onlinecite{moorcroft14}] using the frozen-time stability assumption.

 It must, however, be pointed out that Moorcroft and Fielding [\onlinecite{moorcroft14}] did flag a caveat regarding the frozen-time stability analysis and, therefore, also carried out a numerical integration of the linearized governing equations. However, the authors concluded that results from their frozen-time analysis agreed well with linearized evolution of perturbations for nRP model. Ianniruberto and Marucci [\onlinecite{ianniruberto2017shear}] carried out both linear and non-linear stability analyses using the integral Doi-Edwards model. They reported that no transient shear banding is present for shear rates to the left of the maximum of the constitutive curve, even in the presence of a stress overshoot (also showed for nRP model using frozen time analysis by Moorcroft and Fielding [\onlinecite{moorcroft14}]). For shear rates to the right of the minimum of the constitutive curve, the authors found that while the system is linearly stable, it can be unstable to finite amplitude disturbances, leading to metastability. The time evolution of linearized perturbation and velocity profile evolution for shear start-up to a rather small range of shear rates in the quasi-plateau region of the full monotonic flow curve showed that the time duration of decrease of stress with time is most sensitive to perturbations, and may lead to transient shear banding. They also showed that a decrease in solvent viscosity leads to increase in amplification of perturbations because of shifting of the minimum of the constitutive curve to higher shear rates.

The transient shear banding criterion [\onlinecite{moorcroft14}] for shear start-up flow in a planar Couette geometry appears to be consistent with with the molecular simulation studies of Cao and Likhtman [\onlinecite{cao2012shear}] and Mohagheghi and Khomami [\onlinecite{mohagheghi2016elucidating}]. Zhou \textit{et al.} [\onlinecite{zhou2008modeling}] employed a partially extending and convecting (PEC) model [\onlinecite{larson1984constitutive}] and predicted elastic recoil during shear startup, which is consistent with the criterion for transient shear banding. The authors carried out their analysis in the inertialess limit for a cylindrical Couette geometry, and for start up to shear rates in the non-monotonic region of the constitutive curve. Studies using the VCM model for shear start-up flows in a curvilinear geometry, to a shear rate in the non-monotonic region of the flow curve, reported transient negative velocity profiles both in presence and absence of inertia and also show a correlation with presence of overshoot in shear stress or first normal stress difference [\onlinecite{zhou2012multiple,zhou2014wormlike}]. An increase in inertia diminished the intensity of elastic recoil. The recent work of Ianniruberto and Marrucci [\onlinecite{ianniruberto2017shear}] analyzed both steady-state and transient banding using the integral version of the Doi-Edwards model (with and without convective-constraint release; CCR), and found that their predictions show both kind of results i.e., agreement and disagreement with the findings of Moorcroft and Fielding [\onlinecite{moorcroft14}], as discussed in the following section. Recently, Benzi and coworkers [\onlinecite{benzi2021continuum}] studied a continuum model for yield-stress fluids and showed that transient shear banding is not governed by the overshoot in shear stress. Experimental results of shear startup flow of wormlike micellar solutions and polymeric solutions using planar Couette, cylindrical Couette and cone and plate geometry, primarily by Wang and coworkers, apparently showed a correlation between shear stress overshoot and transient shear banding [\onlinecite{tapadia2006direct,ravindranath2008banding,wang2011homogeneous,boukany2008use,boukany2009exploring,boukany2009shear,hu2007constitutive,hu2008comparison}]. However, Li and coworkers [\onlinecite{li2013flow}] did not observe transient shear banding for a mixture of cis-1,4 polybutadienes and phenyl terminated oligomeric polybutadienes which was later reported to be consistent with results of Wang and coworkers [\onlinecite{wang2014letter}]. A recent experimental effort by Briole et al. [\onlinecite{briole2021shear}] has specifically analyzed the criterion for transient shear banding during shear startup flow proposed by Moorcroft and Fielding [\onlinecite{moorcroft14}, using shear rates in the non-monotonic region of the flow curve. The authors used an aqueous solution of cetyltrimethyl ammonium bromide and sodium nitrate for the case when the ultimate steady state is banded. While Briole et al. noted that stress overshoot may play a crucial role in the nucleation of instability, their results (Fig. 9 of their paper) clearly show that significant departure from the linear velocity profile initiates only at the times near the end of stress overshoot relaxation, even for the case of steady-state shear banding. Thus, in our view, more experiments are needed to ascertain the significance of transient shear banding, especially in systems wherein the eventual steady-state is homogeneous.


An important pre-requisite for the theoretical prediction of transient banding appears to be the presence of a nearly-flat portion in the constitutive curve [\onlinecite{adams2009nonmonotonic,moorcroft14,adams2011transient,moorcroft2013criteria}]. It has been reported that a positive transient eigenvalue is predicted only if shear rate is in the near-flat portion of the constitutive curve. The flatness of the constitutive curve, in turn, arises due to the decrease in the solvent viscosity ratio parameter for most of the viscoelastic constitutive models. In the case of Rolie-Poly model, CCR and entanglement number governs the flatness of the constitutive curve. The solvent viscosity parameter is known to have its own effect in causing a divergent growth rate in the limit of solvent viscosity ratio tending to zero [\onlinecite{tomar2007electrohydrodynamic}]. Earlier work on instabilities in viscoelastic flows with a free surface [\onlinecite{aitken1993rayleigh,tomar2006instability,tomar2007electrohydrodynamic}] have pointed out that in the absence of fluid inertia, the growth rate of instabilities diverge in the limit of zero solvent viscosity. The neglect of inertia, in the case of concentrated polymer solutions and melts, is usually justified owing to the very large viscosities of these systems. However, this is consistent only if the rate of change of fluid velocities remains moderate; when the growth rate diverges, the neglect of fluid inertia becomes untenable. The above-mentioned studies have demonstrated how the growth remains bounded upon inclusion of fluid inertia in the context of instabilities in free-surface flows of viscoelastic fluids. We show later in this manuscript that the significantly high growth rate from frozen-time analysis interpreted as a signature of transient shear banding in the literature [\onlinecite{adams2009nonmonotonic,moorcroft14,adams2011transient,moorcroft2013criteria}] could be sensitive to the the small solvent viscosity values considered in those studies. We then demonstrate the importance of fluid inertia in that limit for all the three constitutive models considered here.
To this end, we also study Giesekus model along with Johnson-Segalman and non-stretching Rolie Poly models to explicitly understand the effect of increasing the flatness of constitutive curve (by decreasing solvent viscosity ratio) on the signatures of transient shear banding. The linearized evolution of perturbations in the case of Giesekus model can provide an independent corroboration of the relation between flatness of the constitutive curve and growth and decay of perturbations, as the Giesekus model does not show any transiently positive eigenvalue (instability within the frozen-time analysis) [\onlinecite{moorcroft14}].

It is worth noting that, although the derivation of the criterion for transient banding during shear start-up has the above-mentioned limitations, the criterion developed by Moorcroft and Fielding [\onlinecite{moorcroft14}] for transient shear banding (within the inertialess limit) during a step increase in stress appears to be devoid of these limitations. This criterion has also been verified experimentally and numerically in literature [\onlinecite{boukany2008use,hu2008comparison,boukany2009shear,boukany2009exploring,hu2005kinetics,divoux2011stress,gibaud2010heterogeneous,grenard2014timescales,sentjabrskaja2015creep,chaudhuri2013onset}].

Therefore, on the basis of above discussion, we identify at least three avenues for improvement, in the transient shear banding study (and criterion) of  Moorcroft and Fielding [\onlinecite{moorcroft14}] during shear start-up: (i) the frozen-time approximation (this has also been noted by Ref. [\onlinecite{moorcroft14}]), (ii) neglect of inertia, and (iii) weaknesses of constitutive equations, especially in the limit of zero solvent viscosity. In the present report, we will provide resolutions to the challenges in (i) and (ii), and we will provide a more comprehensive exploration for the implications of small (or zero) solvent viscosity in the nRP, JS, and Giesekus models.

While earlier numerical studies (Refs. [\onlinecite{moorcroft14,adams2011transient,moorcroft2013criteria}]) have explored the Giesekus and Rolie-Poly models (using frozen-time analysis, linearized evolution of perturbations and full non-linear simulations), the JS model has not been employed hitherto to understand transient banding for shear start-up flow with shear rate in the monotonic region of the constitutive curve. This is particularly intriguing, since the JS model (augmented by stress diffusion) is the canonical model that has been extensively used to understand \textit{steady-state} shear banding. Both the JS and nRP models show a limiting stress strain curve, and therefore, are susceptible to an elastic instability during shear start-up flow as per the criterion of Moorcroft and Fielding [\onlinecite{moorcroft14}]. (The limiting stress strain curve condition is based on the definition of Cauchy elastic material, which is defined as a material in which stress in the material at any point of time depends only on the amount of deformation (strain) present at that point of time [\onlinecite{ogden1997non}].) On the other hand, the Giesekus model does not show a limiting stress strain curve [\onlinecite{moorcroft14}] and therefore, provides a different perspective towards understanding transient shear banding in shear start-up flows using linearized evolution of perturbations. Therefore, we study these three constitutive models for start-up shear flow, especially to shear rates in the monotonic region of the constitutive curves. As discussed above, we employ two different approaches to assess the existence of transient shear banding: a `frozen-time' linear stability analysis which examines whether there is an exponential growth at a given time, and the fundamental matrix approach [\onlinecite{alam1997influence,schmid1994optimal,schmid1994transient}] which characterizes the actual (non-exponential) growth for time-evolving base states, where the perturbations co-evolve with the base-state stresses.

The rest of this paper is structured as follows: In section~\ref{section_model}, we discuss the underlying governing equations and the Johnson-Segalman, non-stretching Rolie-Poly and Giesekus models used. We then discuss the linear stability analysis procedure used in this work. The results are discussed under four categories in Section \ref{section_results}: results from frozen-time analysis, linearized evolution of perturbations, effect of flatness of constitutive curve and effect of inertia. 
The salient conclusions are discussed in section \ref{section_conclusions}.

\section{Models and governing equations} \label{section_model}

We study shear start-up flow of viscoelastic fluids modelled using the Johnson-Segalman [\onlinecite{johnson1977model}], non-stretching Rolie-Poly [\onlinecite{doi1988theory,likhtman2003simple}] and Giesekus models [\onlinecite{giesekus1982simple}]. We consider an incompressible fluid in between two parallel plates which are separated by a distance $H$ in the $y^*$ direction. The plates are infinite in the $x^*$ and $z^*$ directions. At time $t^*=0$ the top plate is set into motion in $x^*$ direction at a constant velocity $U$. Both the plates are initially at rest unless specified and the no-slip boundary condition is applicable at the two plates. In this manuscript, all the variables with superscript $*$ are dimensional quantities, and the ones without a superscript are dimensionless. The continuity equation for this system gets satisfied by itself for the unidirectional flow, \( \nabla \cdot {{\underset{\sim}{u}}^{*}}=0 \), where \({\underset{\sim}{u}}^*\) is the velocity vector. The Cauchy momentum equation can be simplified as follows:
\begin{equation}\label{inertialess}
    \nabla \cdot {\underset{\approx}{\Sigma }^{*}}=\rho\frac{\partial \underset{\sim}u^{*}}{\partial t^{*}},
\end{equation}
where, \({\underset{\approx}{\Sigma }^{*}}\) is the total stress tensor. We initially consider the creeping-flow approximation, but consider the role of inertia later in Sec.\ref{inertia_effect}. The total stress in the fluid is a sum of viscoelastic and Newtonian solvent stresses:

\begin{equation}\label{totalstress}
    {\underset{\approx}{\Sigma }^{*}}={\underset{\approx}{\sigma }^{*}}+{{\mu }_{s}}{\underset{\approx}{\dot{\gamma }}^{*},}
\end{equation}
where, ${\mu }_{s}$ is the solvent viscosity and  ${\underset{\approx}{\dot{\gamma }}}^{*} = (\nabla{{\underset{\sim}{u}^{*}}} + (\nabla{{\underset{\sim}{u}^{*}}})^{T}) $, is the rate-of-strain tensor. We next discuss the three models used in this manuscript. 
  
Stress diffusion is not considered in the linear stability analysis. The following non-dimensional scheme is used for all the three models. Various components of the stress tensor are non-dimensionalized by $\left(\displaystyle \frac{\mu_s+\mu_p}{\tau}\right)$, time by $\tau$, and  velocity by $U$. Here, $\mu_p$ is polymer contribution to solution viscosity and $\tau$ is the relaxation time, depending on the model ($\tau$ = $\tau_D$ is the reptation time in the Rolie-Poly model, and $\tau$ = $\lambda$ is the relaxation time in the Johson-Segalman and Giesekus model). The dimensionless numbers relevant here are the Reynolds number $Re=\frac{\rho U H}{\mu_s+\mu_p}$, Weissenberg number $Wi = \displaystyle \frac{\tau U}{H}$, and ratio of viscosity of solvent to solution viscosity, $\eta_s = \displaystyle \frac{\mu_s}{\mu_s+\mu_p}$.

\textit{Johnson-Segalman model} - Johnson and Segalman developed a phenomenological model for polymer melts by replacing the upper convected derivative in the upper-convected Maxwell (UCM) model with the Gordon-Schowalter derivative [\onlinecite{birddynamics,gordon1972anisotropic,johnson1977model}]. The Gordon-Schowalter derivative introduces a non-affine motion through the slip parameter \( \left( \xi \right) \). For \(  \xi=0 \), the Johnson-Segaman (hereafter JS) model, with Newtonian solvent, reduces to the Oldroyd-B model, while for \(  \xi=2 \) it reduces to the Oldroyd-A model [\onlinecite{birddynamics}]. The constitutive equation of the JS model is given as follows
\begin{equation}
 \underset{\approx}\sigma^*+\lambda \overset{\square}{\underset{\approx}{\sigma}^*}  ={\mu}_p \underset{\approx}{\dot{\gamma}^*},
\end{equation}
where, $\underset{\approx}\sigma^*$ is the viscoelastic stress tensor, $\lambda$ is the relaxation time of the polymeric solution, $\mu_p$ is the polymer contribution to the solution viscosity. The equilibrium stress tensor for the JS model is \(\underset{\approx}{\sigma_0} =\underset{\approx}{0}\). The Gordon-Schowalter derivative $ \overset{\square}{\underset{\approx}{\sigma}}$ is expressed as follows:
\begin{equation}
   \overset{\square}{\underset{\approx}{\sigma^*}}= \left( 1-\frac{\xi}{2} \right){\overset{\triangledown}{\underset{\approx}{\sigma^*}}}+\left( \frac{\xi}{2} \right){\overset{\triangle}{\underset{\approx}{\sigma^*}}},
\end{equation}
where, ${\overset{\triangledown}{\underset{\approx}{\sigma^*}}}$ is the upper convected derivative and ${\overset{\triangle}{\underset{\approx}{\sigma^*}}}$ is the lower convected derivative [\onlinecite{larson2013constitutive}]. The simplified non-dimensional component-wise equations of the model for simple shear flow are given below:
\begin{equation}\label{jsxy}
    \frac{\partial {{\sigma }_{xy}}}{\partial t} = \left( -\left( \frac{\xi }{2} \right){{\sigma }_{xx}}+\left( 1-\frac{\xi }{2} \right){{\sigma }_{yy}}+\left( 1-\eta_s  \right) \right)\dot{\gamma }Wi-{{\sigma }_{xy}},
\end{equation}
\begin{equation}\label{jsxx}
\frac{\partial {{\sigma }_{xx}}}{\partial t} = 2\left( \left( 1-\frac{\xi }{2} \right){{\sigma }_{xy}} \right)\dot{\gamma }Wi-{{\sigma }_{xx}},
\end{equation}
and
\begin{equation}\label{jsyy}
\frac{\partial {{\sigma }_{yy}}}{\partial t} = -\left( \xi {{\sigma }_{xy}} \right)\dot{\gamma }Wi-{{\sigma }_{yy}}.
\end{equation}
All the results for JS model in this manuscript are obtained for $\xi=0.01$.


\textit{Rolie-Poly model} - Likhtman and Graham presented a simplified single-mode molecular model from the tube model [\onlinecite{doi1988theory,likhtman2003simple}]. This model was termed by Likhtman and Graham as Rolie-Poly (ROuse LInear Entangled POLYmers) model. The three kinds of relaxation mechanisms considered in the model are reptation, retraction and convective constraint release (CCR). The viscoelastic stress is given by the following equation 
\begin{equation} \label{rpmodel}
    \left( \underset{\approx}{\sigma}^*-\underset{\approx}{I} \right)+{\tau _D}\overset{\nabla }{\mathop{\underset{\approx}{\sigma}^*}}=-2\frac{{{\tau }_{D}}}{{{\tau }_{R}}}\left( 1-\sqrt{\frac{3}{tr\left( \underset{\approx}{\sigma}^* \right)}} \right)\left( \underset{\approx}{\sigma}^*+\beta \left( \underset{\approx}{\sigma}^*-\underset{\approx}{I}  \right){{\left(\sqrt{ \frac{3}{tr\left( \underset{\approx}{\sigma}^* \right)}} \right)}^{-2\delta }} \right),
\end{equation}
where, \(\underset{\approx}{\sigma}^*\) is the viscoelastic stress tensor, \( {\beta} \)  fixes the CCR effectiveness, \( {\delta} \) defines the strength of CCR because of chain stretch mechanism, its value is fixed throughout the manuscript as 1/2 following [\onlinecite{likhtman2003simple,holroyd2017analytic}] and $\underset{\approx}{I}$ is the identity tensor. Here, \({\tau }_{D} \) is a relaxation time associated with reptation mechanism and ${\tau }_{R}$ is a relaxation time for chain stretch mechanism. Here the two relaxation times are fixed by the number of entanglements \(Z\), \( \left(Z=\displaystyle \frac{{\tau }_{D}}{{3\tau }_{R}}\right)\). The equilibrium value of stress tensor is such that at rest \(\underset{\approx}{\sigma_0} =\underset{\approx}{I}\).

Likhtman and Graham [\onlinecite{likhtman2003simple}] also discussed the case of \({\tau }_{R}\rightarrow 0\) and consequently \(tr\left(\underset{\approx}{\sigma}^*\right)=3+\Delta\) (and \(\Delta = 0\) because \({\tau }_{R}\rightarrow 0\)), the Rolie-Poly model can be further simplified as given below:
\begin{equation} \label{nrpmodel}
\left( \underset{\approx}{\sigma}^*-\underset{\approx}{I} \right)+{\tau _D}\overset{\nabla }{\mathop{\underset{\approx}{\sigma}^*}}=-\frac{2}{3}{\tau }_{D}\left(tr\left(\underset{\sim}{\nabla}^* \underset{\sim}{u}^*\cdot\underset{\approx}{\sigma}^*\right)\right)\left( \underset{\approx}{\sigma}^*+\beta \left( \underset{\approx}{\sigma}^*-\underset{\approx}{I}  \right)\right).
\end{equation}
The  above equation is known as non-stretching Rolie-Poly (hereafter nRP) model. The simplified non-dimensional component-wise equations of nRP model for the simple shear flow are as follows: 
\begin{equation}\label{nrpxy}
    \frac{\partial {{\sigma}_{xy}}}{\partial t}=Wi\dot{\gamma }\left( {{\sigma}_{yy}}-\frac{2}{3}\sigma_{xy}^{2}\left( 1+\beta  \right) \right)-{{\sigma}_{xy}},
\end{equation}
and
\begin{equation}\label{nrpxx}
    \frac{\partial {{\sigma}_{yy}}}{\partial t}=\frac{2}{3}Wi\dot{\gamma }\left( \beta {{\sigma}_{xy}}-{{\sigma}_{yy}}\sigma_{xy}^{{}}\left( 1+\beta  \right) \right)-\left( {{\sigma}_{yy}}-1 \right).
\end{equation}
All the studies for nRP model in this manuscript are obtained for $\beta=1$ unless otherwise mentioned.

\textit{Giesekus model} - A non-linear phenomenological model proposed by Giesekus [\onlinecite{giesekus1982simple}] has been used extensively for fitting transient and steady-state data of polymeric solution [\onlinecite{helgeson2009relating,khair2016large,vlassopoulos1995generalized}]. The presence of the extra quadratic term of stress leads to prediction of shear thinning and first and second normal stress differences quite accurately by accounting for anisotropic drag experienced by polymer molecules. The constitutive model in dimensional form is expressed as:
  \begin{equation} \label{gies_model}
     \underset{\approx}{\sigma}^* +{\lambda}\overset{\nabla }{\mathop{\underset{\approx}{\sigma}^*}}=-\alpha\frac{\lambda}{\mu_p}(\underset{\approx}{\sigma}^*\cdot\underset{\approx}{\sigma}^*)+\mu_{p}\underset{\approx}{\dot{\gamma}^*},
\end{equation}
where $\alpha$ is a drag anisotropy parameter. All the results for Giesekus model in this manuscript are obtained for $\alpha=0.1$. The simplified non-dimensional component-wise equations of Giesekus model for simple shear flow are as follows:
\begin{equation}\label{gxy}
    \frac{\partial \sigma_{xy}}{\partial t}=-\frac{\alpha}{(1-\eta_s)}(\sigma_{xx}+\sigma_{yy})\sigma_{xy}+[(1-\eta_s)+\sigma_{yy}]Wi\dot{\gamma}-\sigma_{xy},
\end{equation}
\begin{equation}\label{gxx}
    \frac{\partial \sigma_{xx}}{\partial t}=-\frac{\alpha}{(1-\eta_s)}(\sigma_{xx}^2+\sigma_{xy}^2)+2Wi\dot{\gamma}\sigma_{xy}-\sigma_{xx},
\end{equation}
and
\begin{equation}\label{gyy}
    \frac{\partial \sigma_{yy}}{\partial t}=-\frac{\alpha}{(1-\eta_s)}(\sigma_{yy}^2+\sigma_{xy}^2)-\sigma_{yy}.
\end{equation}

For the above three models, we utilize linear stability analysis to understand the effect of any perturbation on the flow as discussed below.

\subsection{Linear stability analysis}
\label{lsa}
\subsubsection{Base state} The base state whose linear stability is examined in this work is not a steady one, but evolves with time from the rest state to the eventual steady state. During the evolution of base state, we consider flow in the inertia-less limit. Upon imposition of a step shear rate, the fluid velocity profile instantaneously acquires the linear velocity distribution. However, the stresses evolve as a function of time before reaching the steady-state values, owing to the elastic nature of the fluids considered.  

\subsubsection{Linearized governing equations}

At the outset, we emphasize that the linearized analyses carried out in this work are capable of predicting only the \textit{onset} of transient shear banding due to \textit{infinitesimal} disturbances. There are two possible outcomes:  (1) if the flow is linearly unstable, then the linearized analysis predicts that the given state is unstable, but does not provide any information about the eventual state the system will evolve into; that information can be obtained only by a nonlinear simulation of the full governing equations, (2) if the flow is linearly stable, there is still the possibility of it becoming unstable to finite amplitude disturbances, which can again be investigated only by nonlinear simulations.  While the focus of the present work is on linear stability, the nonlinear dynamics of shear start-up will be discussed in a future communication [\onlinecite{sharma2022transient}].

We perform a linear stability analysis of the above-mentioned time-dependent base state for all the three models. We neglect the diffusion terms in these three models before linearisation, following Moorcroft and Fielding [\onlinecite{moorcroft14}]. Here, we impose a perturbation (terms with superscript $n$, ${{\hat{\underset{\approx}{\sigma}}^{n}}}(t)$ and ${{\hat{\dot{\underset{\approx}{\gamma}}}}}^{n}(t)$ ( ${{\hat{{\underset{\sim}{u}}}}}^{n}(t)$ for non-zero inertial case of linearized evolution of perturbations), where, $n$ is the index for mode number) initially and obtain their time evolution along with the base state to determine the stability of the system at a given time. The perturbations are substituted into the base state equations to obtain their evolution according to the following equations:

\begin{equation}\label{sigma0}
    \underset{\approx}{\sigma}(t)={\underset{\approx}{\sigma}^{0}}(t)+\sum\nolimits_{n=1}^{\infty }{{\hat{\underset{\approx}{\sigma}}^{n}}}(t)\cos(n\pi y),
\end{equation}
\begin{equation}\label{gamma0}
  \underset{\approx}{\dot{\gamma}}(t)={\underset{\approx}{\dot{\gamma}}^{0}}(t)+\sum\nolimits_{n=1}^{\infty }{{\hat{\dot{\underset{\approx}{\gamma}}}}}^{n}(t)\cos(n\pi y),
\end{equation}
and
\begin{equation}\label{u0}
  \underset{\sim}{u}(t)={ \underset{\sim}{u}^{0}}(t)+\sum\nolimits_{n=1}^{\infty }{{\hat{ \underset{\sim}{u}}}}^{n}(t)\sin(n\pi y).
\end{equation}
The cosine terms for spatial perturbations are used in order to satisfy no slip boundary conditions. (The cosine spatial perturbations for shear rate are obtained by differentiating sinusoidal spatial perturbation for velocity to satisfy boundary condition.) We substitute Eqs.~\ref{sigma0}-\ref{u0} into the Cauchy momentum equation (Eq.~\ref{inertialess}) and obtain following linearized equation:
\begin{equation}\label{cauchy_momentum_pert_inertia}
    Re\frac{d \hat{u^n}}{d t}=-n^2 \pi^2 (\hat{\sigma }_{xy}^{n}+\eta_sWi\hat{u^n}),
\end{equation}
and for $Re=0$, the above equation simplifies to
\begin{equation}\label{Sigma0}
    {{\Sigma }_{xy}}\left( t \right)=\Sigma _{xy}^{0}\left( t \right),
\end{equation}
and
\begin{equation}\label{cauchy_momentum_pert}
       {{\hat{\dot{\gamma}}}}^{n}=-\frac{\hat{\sigma }_{xy}^{n}}{\eta_{s}Wi}.
\end{equation}
As mentioned above, the evolution of perturbation is calculated by substituting Eqs.~\ref{sigma0}-\ref{gamma0} and \ref{Sigma0}, for inertialess limit, and Eq.~\ref{sigma0}, \ref{u0} and \ref{cauchy_momentum_pert_inertia} for flow in presence of inertia, into Eqs.~\ref{jsxy}-\ref{jsyy} (for JS model), Eqs.~\ref{nrpxy}-\ref{nrpxx} (for nRP model), and  Eqs.~\ref{gxy}-\ref{gyy} (for Giesekus model). As a result, we obtain a set of linearized ordinary differential equations (ODEs) after neglecting the non linear terms. The ordinary differential equations (ODEs) pertaining to the JS model are

\begin{equation}\label{jsxyp}
    \frac{d \hat{\sigma }_{xy}^{n}}{d t}=\left\{ \left( {{{\dot{\gamma }}}^{0}}\hat{\sigma }_{yy}^{n}+{{{\hat{\dot{\gamma }}}}^{n}}\sigma _{yy}^{0} \right)\left( 1-\frac{\xi }{2} \right)-\left( {{{\dot{\gamma }}}^{0}}\hat{\sigma }_{xx}^{n}+{{{\hat{\dot{\gamma }}}}^{n}}\sigma _{xx}^{0} \right)\frac{\xi }{2}+\left( 1-\eta_s  \right){{{\hat{\dot{\gamma }}}}^{n}} \right\}Wi-\hat{\sigma }_{xy}^{n},
\end{equation}
\begin{equation}\label{jsxxp}
    \frac{d \hat{\sigma }_{xx}^{n}}{d t}=2\left( {{{\dot{\gamma }}}^{0}}\hat{\sigma }_{xy}^{n}+{{{\hat{\dot{\gamma }}}}^{n}}\sigma _{xy}^{0} \right)\left( 1-\frac{\xi }{2} \right)Wi-\hat{\sigma }_{xx}^{n},
\end{equation}
and
\begin{equation}\label{jsyyp}
    \frac{d \hat{\sigma }_{yy}^{n}}{d t}=-\left( {{{\dot{\gamma }}}^{0}}\hat{\sigma }_{xy}^{n}+{{{\hat{\dot{\gamma }}}}^{n}}\sigma _{xy}^{0} \right)\xi Wi-\hat{\sigma }_{yy}^{n}.
\end{equation}
The ordinary differential equations (ODEs) pertaining to the nRP model are
\begin{equation}\label{nrpxyp}
    \frac{d \hat{\sigma }_{xy}^{n}}{d t}=Wi\left( \left( {{{\dot{\gamma }}}_{0}}\hat{\sigma }_{yy}^{n}+{{{\hat{\dot{\gamma }}}}_{n}}\sigma_{yy}^{0} \right)-\frac{2}{3}\left( 2{{{\dot{\gamma }}}_{0}}\sigma_{xy}^{0}\hat{\sigma }_{xy}^{n}+{{{\hat{\dot{\gamma }}}}_{n}}\sigma{{_{xy}^{0}}^{2}} \right)\left( 1+\beta  \right) \right)-\hat{\sigma }_{xy}^{n},
\end{equation}
and
\begin{equation}\label{nrpxxp}
    \frac{d \hat{\sigma }_{yy}^{n}}{d t}=-\frac{2}{3}Wi\left( \left( {{{\dot{\gamma }}}_{0}}\sigma_{xy}^{0}\hat{\sigma }_{yy}^{n}+{{{\dot{\gamma }}}_{0}}\sigma_{yy}^{0}\hat{\sigma }_{xy}^{n}+{{{\hat{\dot{\gamma }}}}_{n}}\sigma_{xy}^{0}\sigma_{yy}^{0} \right)\left( 1+\beta  \right)-\left( {{{\dot{\gamma }}}_{0}}\hat{\sigma }_{xy}^{n}+{{{\hat{\dot{\gamma }}}}_{n}}\sigma_{xy}^{0} \right)\beta  \right)-\hat{\sigma }_{yy}^{n}.
\end{equation}
The ordinary differential equations (ODEs) pertaining to the Giesekus model are
\begin{equation}\label{gxyp}
     \frac{d \hat{\sigma }_{xy}^{n}}{d t}=-\frac{\alpha}{(1-\eta_s)}(\sigma_{xx}^0\sigma_{xy}^n+\sigma_{xx}^n\sigma_{xy}^0+\sigma_{yy}^0\sigma_{xy}^n+\sigma_{yy}^n\sigma_{xy}^0)+((1-\eta_s)Wi\dot{\gamma}^n+\sigma_{yy}^0Wi\dot{\gamma}^n+\sigma_{yy}^nWi\dot{\gamma}^0)-\sigma_{xy}^n,
\end{equation}
\begin{equation}\label{gxxp}
    \frac{d  \hat{\sigma }_{xx}^{n}}{d t}=-\frac{\alpha}{(1-\eta_s)}(2\sigma_{xx}^0\sigma_{xx}^n+2\sigma_{xy}^0\sigma_{xy}^n)+2Wi(\dot{\gamma}^n\sigma_{xy}^0+\dot{\gamma}^0\sigma_{xy}^n)-\sigma_{xx}^n,
\end{equation}
and
\begin{equation}\label{gyyp}
    \frac{d  \hat{\sigma }_{yy}^{n}}{d t}=-\frac{\alpha}{(1-\eta_s)}(2\sigma_{yy}^0\sigma_{yy}^n+2\sigma_{xy}^0\sigma_{xy}^n)-\sigma_{yy}^n.
\end{equation}

We observe from Eqs.~\ref{jsxyp}-\ref{jsyyp}, Eqs.~\ref{nrpxyp}-\ref{nrpxxp} and Eqs.~\ref{gxyp}-\ref{gyyp} that all the cosine terms get cancelled out, and the behavior of all the modes of a perturbation is identical for flow under creeping flow approximation. However, the evolution of perturbation terms, as well as eigenvalues will depend on value of $n$ in presence of inertia (Eq. \ref{cauchy_momentum_pert_inertia}).
The initial condition for the base state is the rest state.

We perform two types of linear stability analysis: 

(i) a `frozen-time' analysis, wherein at each instant of time in the evolution of the base state, the governing equations are linearized, and a modal stability analysis is carried out to determine the eigenvalue, $\omega$, that characterizes the exponential growth or decay of perturbations. Such an analysis would be strictly valid only if the growth rate of perturbations (if the flow is unstable) is large compared to the rate of change of base state quantities. However, it must be noted that it is not possible to guarantee whether the frozen-time analysis will be accurate \textit{a priori}, and the results obtained from the frozen-time analysis must be checked for self-consistency \textit{post-facto}. In the frozen-time analysis, the base state about which the perturbations are linearized is a function of time, the eigenvalues obtained from the same would also be a function of time.

(ii) In the second method, we linearize the governing equations about the time-evolving base state, and solve for the evolution of perturbations numerically. This method does not involve the assumption of exponential growth or decay, and further does not involve the time-freezing of the base-state quantities. The evolution of perturbations for the time varying base state can be represented as:

\begin{equation}\label{perturbation_ode}
    \frac{d\underset{\sim}{x}(t)}{dt}=\underset{\approx}{A}(t)\underset{\sim}{x}(t)
\end{equation}
    where, components of $\underset{\sim}{x}(t)$ are $\sigma_{xy}^n$, $\sigma_{xx}^n$ and $\sigma_{yy}^n$, and $\underset{\approx}{A}(t)$ contains the base state terms which are also a function of time. We follow Refs. [\onlinecite{alam1997influence,strang1997linear,schmid1994optimal,schmid1994transient}] and represent $\underset{\sim}{x}(t)$ using the fundamental matrix $\underset{\approx}{Y}(t)$, as follows:
  \begin{equation}\label{fundamental_matrix}
     \underset{\sim}{x}(t)=\underset{\approx}{Y}(t)\underset{\sim}{x}(0)
 \end{equation}
At $t=0$,~$\underset{\approx}{Y}(0)=\underset{\approx}{I}$. The time-evolution of Y(t) is governed by
\begin{equation}\label{fundamental_perturbation_ode}
    \frac{d\underset{\approx}{Y}(t)}{dt}=\underset{\approx}{A}(t)\underset{\approx}{Y}(t)
\end{equation}

In order to obtain the maximum amplification of perturbations at any time $t$, we determine growth coefficient $G(t)$, which is defined as:

\begin{equation}\label{G_equation}
   G(t)=\lim_{\underset{\sim}{x}(0)\neq0}sup~\frac{|\underset{\sim}{x}(t)|}{|\underset{\sim}{x}(0)|}
\end{equation}

Substituting Eq. \ref{fundamental_matrix} into \ref{G_equation}, we get
\begin{equation}\label{G_final_equation}
    G(t)=\lim_{\underset{\sim}{x}(0)\neq0}sup~\frac{||\underset{\approx}{Y}(t)\underset{\sim}{x}(0)||}{||\underset{\sim}{x}(0)||}
    =||\underset{\approx}{Y}(t)||
\end{equation}

Therefore, the 2-norm of the fundamental matrix $\underset{\approx}{Y}(t)$  is a measure of the maximum amplification of perturbations and is independent of any initial condition of $\underset{\sim}{x}(t)$. We can also determine a growth rate $(M(t))$, which is analogous to the eigenvalue in frozen-time analysis approach, as follows:

\begin{equation}\label{Mt}
   M(t)=\frac{1}{G(t)}\frac{dG(t)}{dt}
\end{equation}

We calculate the variation of $G(t)$ and $M(t)$ to determine the evolution of linearized perturbations during shear start-up, and compare the growth rate $M(t)$ with the eigenvalue rate from frozen-time analysis. The first positive value of $M(t)$ shows the time at which the growth of perturbation gets triggered and the peak of $M(t)$ shows the time at which perturbations have the maximum growth rate.

As evident from the perturbation ODEs, the eigenvalue is a function of base state. The linearized evolution of perturbation growth coefficient, $G(t)$ and growth rate, $M(t)$ is obtained by solving the perturbation ODEs, using the MATLAB\textsuperscript{\textregistered}  \textit{ode}45 inbuilt routine. The results obtained for the JS model are benchmarked with the known analytical solutions available in the limiting cases of Oldroyd-B and Oldroyd-A models. The base state and frozen-time analysis results from the nRP and Giesekus models obtained in our study are benchmarked with the published results of Moorcroft and Fielding [\onlinecite{moorcroft14}]. All the linear stability analysis results are further verified to converge with decrease in time stepping.

\section{Results and discussion} \label{section_results}

\begin{figure}[h]

 \subfigure[]{
    \includegraphics[scale=0.28]{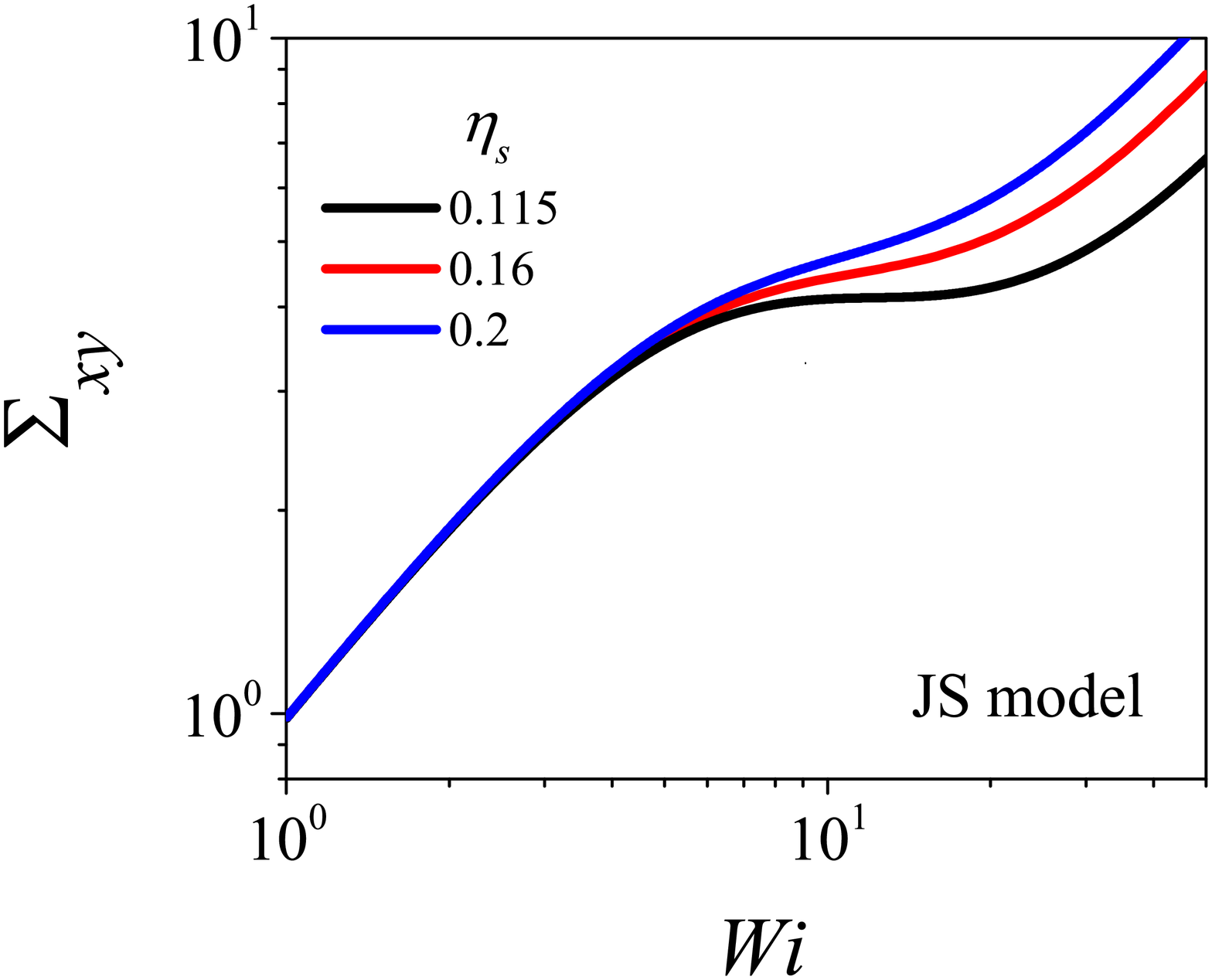}
    \label{fc_js}
  } \subfigure[]{
    \includegraphics[scale=0.28]{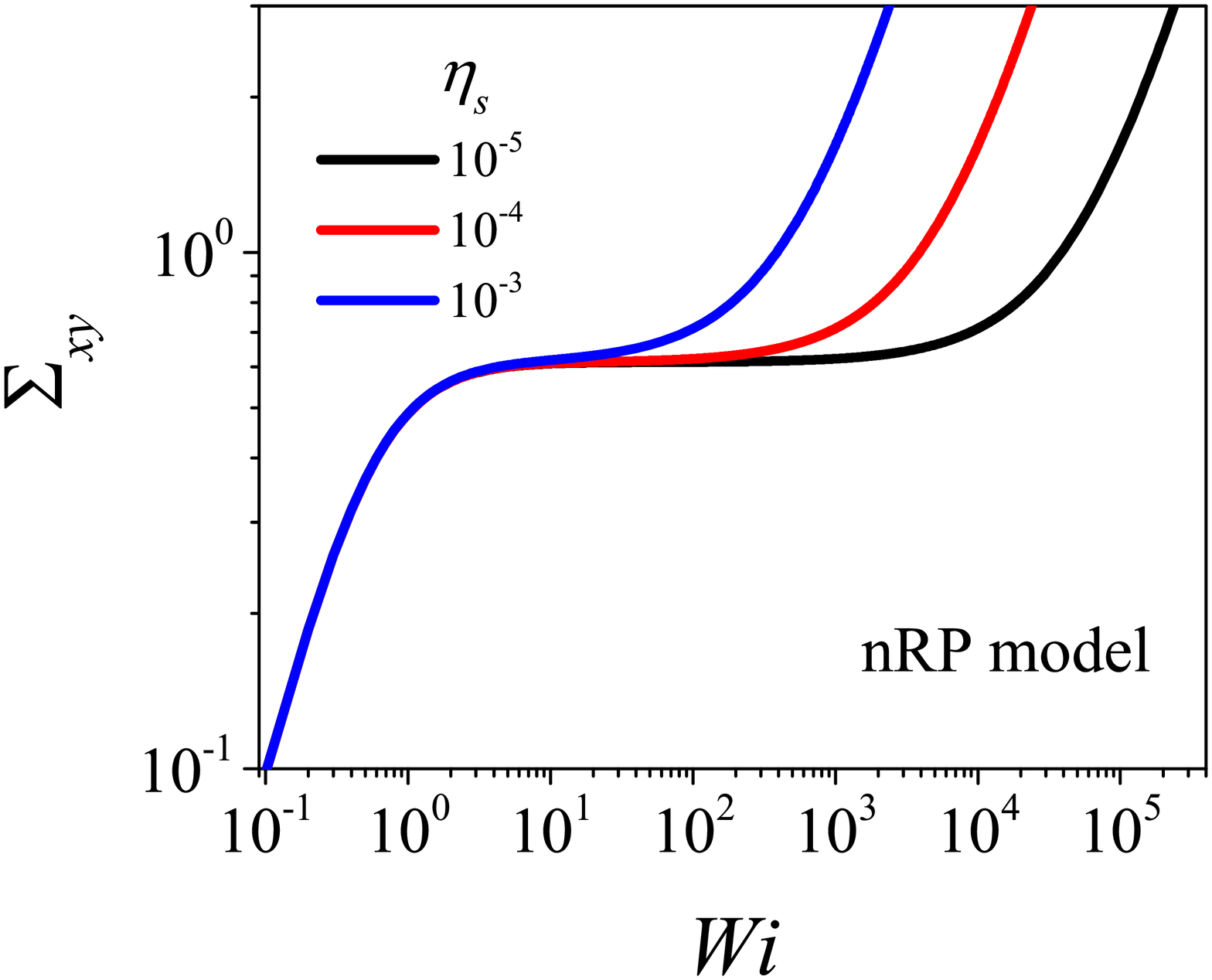}
    \label{fc_nRP}
  }
  \subfigure[]{
    \includegraphics[scale=0.28]{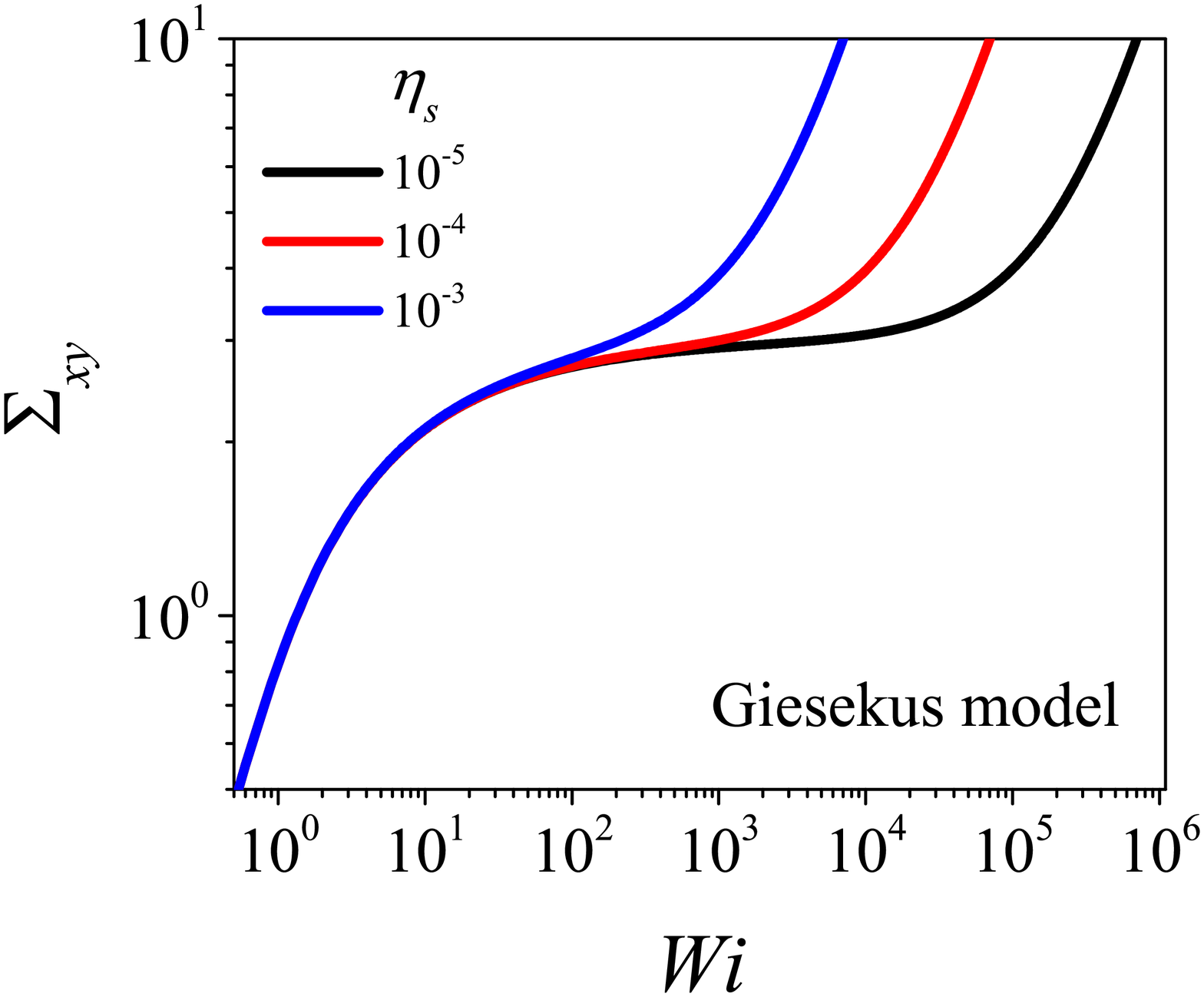}
    \label{fc_giesekus}
  }
   \subfigure[]{
    \includegraphics[scale=0.28]{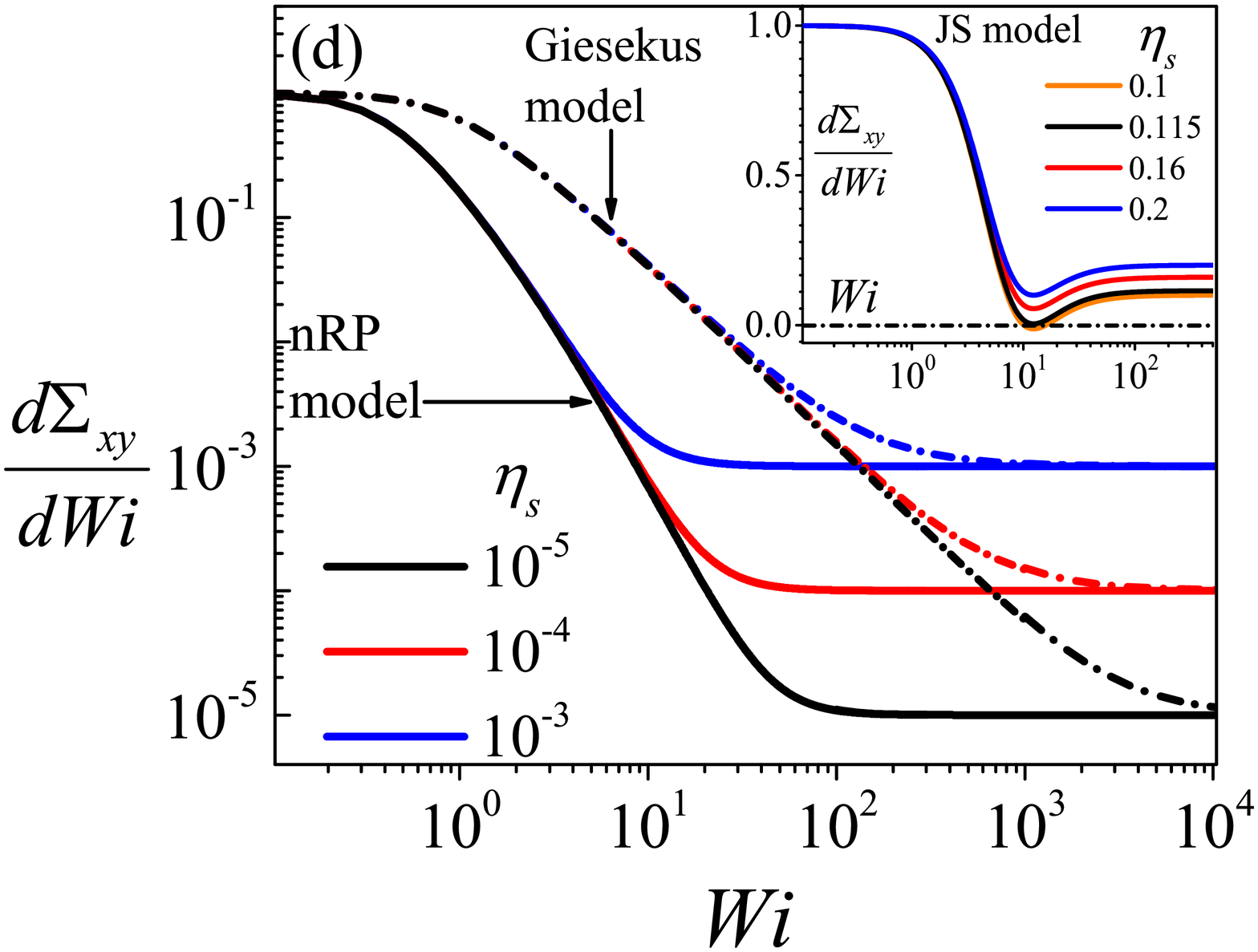}
    \label{ds_dW}
  }
\caption{Steady-state constitutive curves of (a) JS model, (b) nRP model, and (c) Giesekus model shows the steady state relation between total shear stress and $Wi$. In the figures (a), (b) and (c), nearly-flat portion of the constitutive curve of all the three models increases with decrease in solvent viscosity, $\eta_s$. In (a) $\xi=0.01$, in (b) $\beta=1$, and in (c) $\alpha=0.1$ for all three constitutive curves of each model. Figure (d) shows the variation of $\displaystyle \frac{d\Sigma_{xy}}{dWi}$ as a function of $Wi$ for the nRP (solid lines) and Giesekus models (dashed lines). Figure (d) inset shows results for JS model.}
\label{fig:flow_curve}
\end{figure}

We first present the steady-state constitutive curves for the JS, nRP and Giesekus models for different ratios of solvent viscosity to zero shear viscosity of the solution ($\eta_{s}$) in Fig.~\ref{fig:flow_curve}. A common feature in the constitutive curves of all the three models is that upon decreasing $\eta_s$, the flatness of the least steeper portion of the constitutive curve increases, which is also shown in Fig. \ref{ds_dW}. Figure \ref{ds_dW} also shows that compared to nRP and Giesekus models, the solvent viscosity ratio $\eta_s$ cannot be decreased below $1/9$ for the JS model owing to the nonmonotonic nature of the constitutive curve for $\eta_s < 1/9$. This connection between the flatness of the constitutive curve and $\eta_s$ will be helpful in the ensuing discussion of our results.

\begin{figure}[h]
\subfigure[]{
    \includegraphics[scale=0.28]{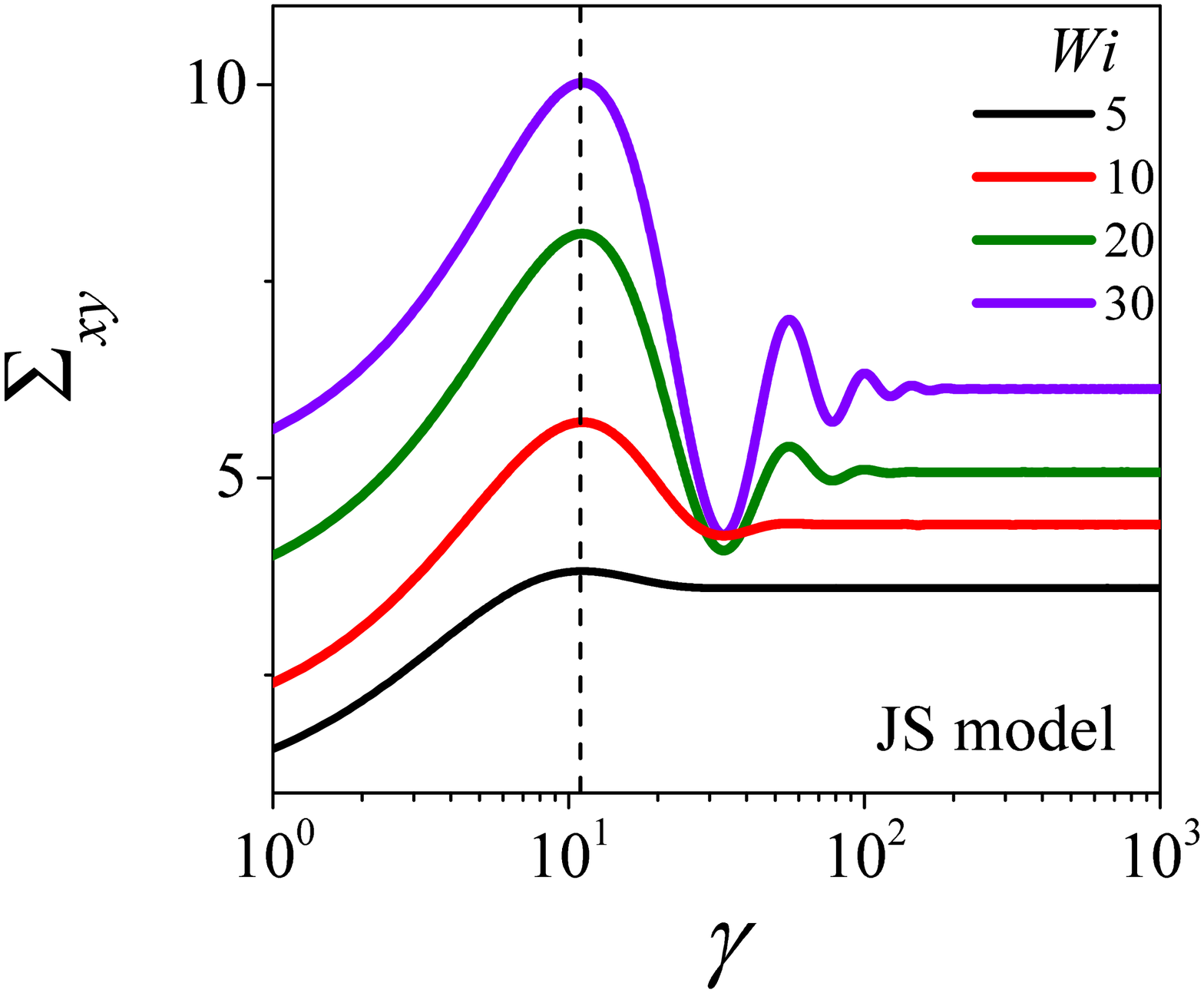}
    \label{lssa_ss_js}
  } \subfigure[]{
    \includegraphics[scale=0.28]{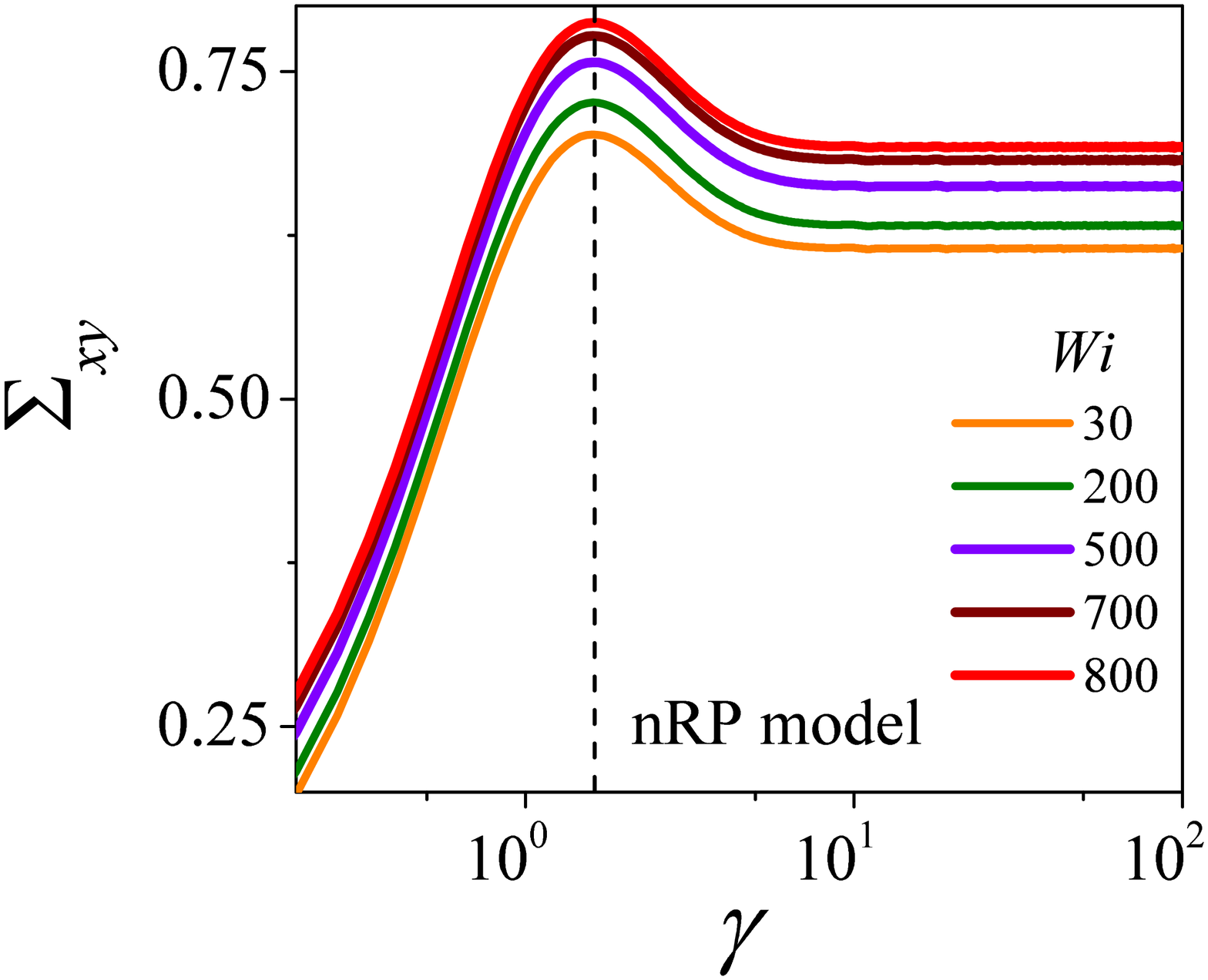}
    \label{lssa_ss_nrp}
  }
  \subfigure[]{
    \includegraphics[scale=0.28]{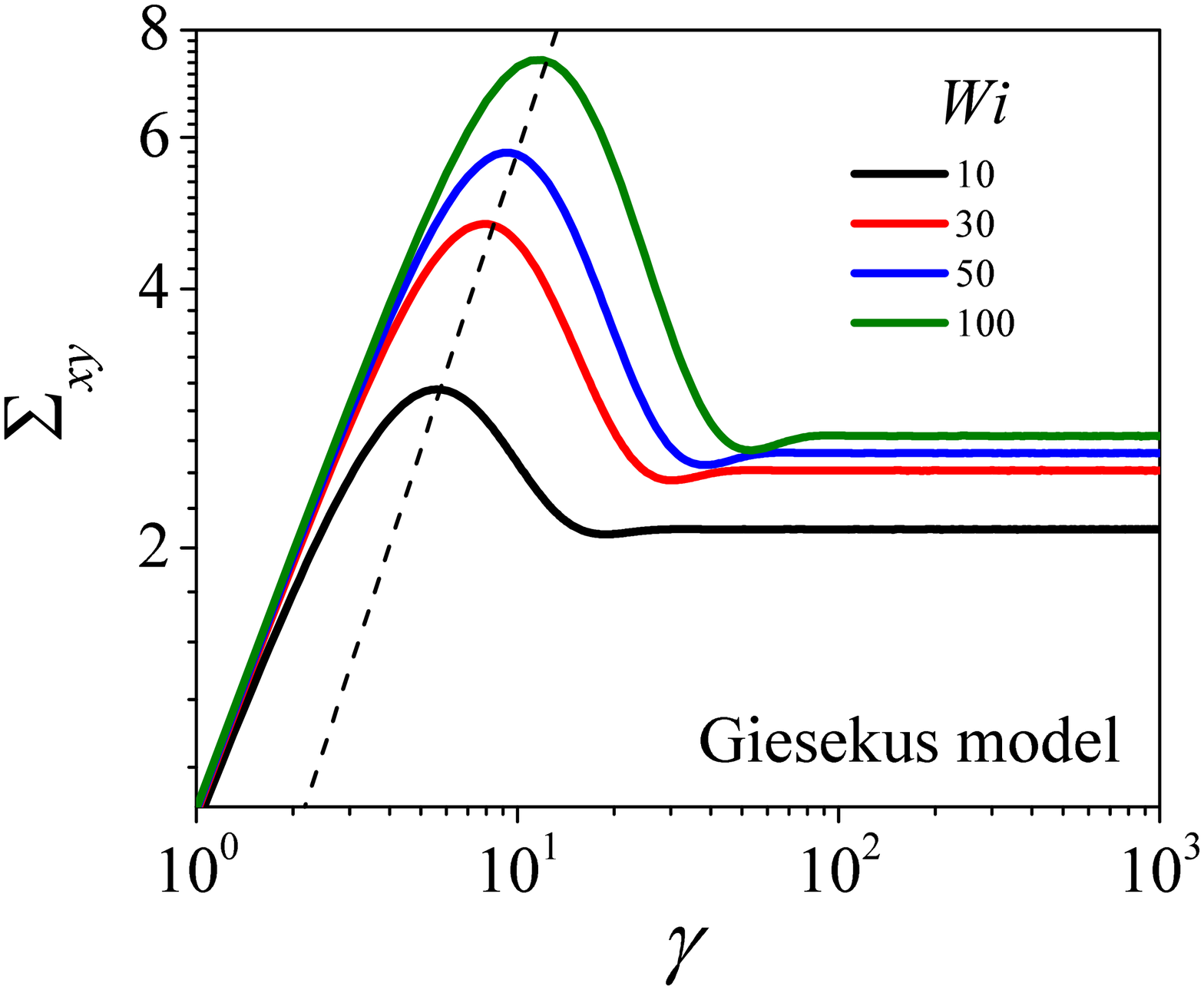}
    \label{lssa_ss_giesekus}
  }

\caption{Shear stress evolution as a function of strain for JS model $(\eta_s=0.16)$, nRP model $(\eta_s=10^{-4})$ and Giesekus model $(\eta_s=10^{-4})$ for shear start-up at different value of shear rate $(Wi)$. Figure (a) and (b) depicts that both JS and nRP model have `limiting stress-strain curve' as shear stress overshoot is at same value of strain irrespective of the value of $Wi$. The Giesekus model does not demonstrate limiting stress strain curve. The dashed line on the three figures shows the locus of strain (constant or varying) corresponding to stress overshoot.}
\label{fig:lsa_lssc}
\end{figure}

We also present the limiting stress-strain curve for JS, nRP and Giesekus models in Fig.~\ref{fig:lsa_lssc}. We reiterate the observation of Moorcroft and Fielding [\onlinecite{moorcroft14}] for nRP and Giesekus models. The nRP model exhibits a stress overshoot that is independent of strain rate as it occurs at the same value of shear strain, and shows a limiting stress-strain curve as shown in Fig. \ref{lssa_ss_nrp}. Figure \ref{lssa_ss_giesekus} shows the case of Giesekus model, in which the shear stress overshoot depends on shear rate and occurs at higher value of shear strain for higher strain rates. We find that, similar to the nRP model, the JS model also exhibits a limiting stress-strain curve, as shown in Fig. \ref{lssa_ss_js}. According to the criterion proposed in literature [\onlinecite{moorcroft14}], figures~\ref{lssa_ss_js} and \ref{lssa_ss_nrp} demonstrate  the capability of both the models to exhibit transient shear banding. 
We compare results for both types of models which do or do not show a limiting stress strain curve, in order to critically evaluate the analytical criterion for transient shear banding proposed in the literature [\onlinecite{moorcroft14}] and analyse various aspects of transient dynamics in the following subsections.

\subsection{Frozen-time analysis}
\begin{figure}[h]
    \includegraphics[scale=0.4]{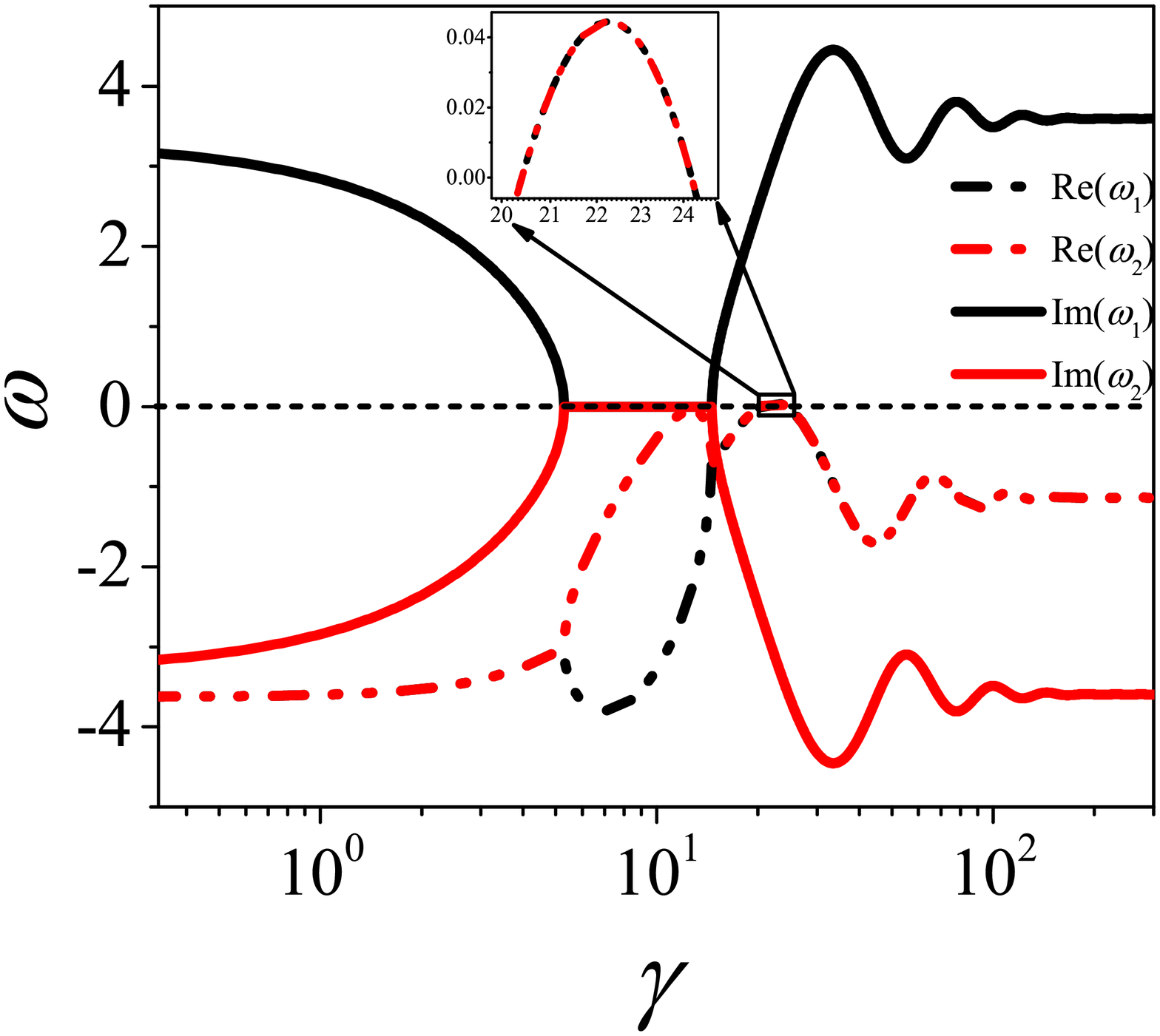}
    
\caption{Evolution of eigenvalues for the Johnson-Segalman model from the frozen-time analysis during shear start-up flow to the monotonic region of the flow curve ($Wi=30$, $\eta_s=0.16$). Real and imaginary part of eigenvalues clearly shows a Hopf bifurcation during their evolution to the steady-state. An additional eigenvalue, which is not shown in this figure, is equal to $-1+0i$ for all $t$. The black short dash line shows the $\omega=0$ line.}
\label{fig:JS_hopf}
\end{figure}
The outcome of the frozen-time stability analysis are the eigenvalue(s) at each time instant, and if the real part of at least one of the eigenvalues is positive, that would then be suggestive of a transient growth of stress fluctuations. The transient eigenvalues of the nRP model, for all possible shear rates and at all times, are observed to be real and distinct, in agreement with the earlier results of Moorcroft and Fielding [\onlinecite{moorcroft14}]. The transient eigenvalues may be positive temporarily (for an intermediate time range) or permanently (for $t \rightarrow \infty$) for nonmonotonic constitutive curves. Our results show that for the nRP model, only one (real) eigenvalue is positive at a given time. The transient eigenvalues of the Giesekus model are observed to have a negative real part and zero or non-zero imaginary part. Hence, the frozen-time eigenvalue analysis for these models appear to be broadly consistent with the results of the eigenvalue analysis of Ref. [\onlinecite{moorcroft14}]. %

Interestingly, results from the JS model show the presence of two complex-conjugate eigenvalues in the transient evolution for shear start-up flow for some values of the shear rate. Furthermore, for a certain time range, the two complex conjugate eigenvalues (with negative real parts) become unstable simultaneously, which is the signature of a Hopf bifurcation [\onlinecite{Drazinreid}]. Figure \ref{fig:JS_hopf} shows the real and imaginary part of the two eigenvalues, $\omega_1$ and $\omega_2$, obtained from the JS model (equations \ref{jsxyp}-\ref{jsyyp}) during shear start-up  to monotonic region of the constitutive curve ($Wi=30$, $\eta_s=0.16$). It must be noted that the third eigenvalue is $\omega_3=-1+0i$ for the complete duration of shear start-up, and hence need not be considered further. At early times $(0 < t < 0.18)$, the two eigenvalues $\omega_1$ and $\omega_2$ are complex conjugates, with the same real part and with equal and opposite imaginary parts. 
The following criterion for the eigenvalues to have a positive real part can be derived from the linearized governing equations of the JS model:

\begin{equation}\label{js_hopf_criteria}
  \sigma_{xx}^{0}\xi+\sigma_{yy}^{0}(\xi-2) > 2(1+\eta_s).  
\end{equation}
 The eigenvalues are complex if:
\begin{equation}\label{eigen_complex}
   (\sigma_{yy}^{0}(\xi-2)+\sigma_{xx}^{0}\xi-2)^2+4\eta_s(\sigma_{yy}^{0}(\xi-2)+\sigma_{xx}^{0}\xi-4\sigma_{xy}^{0}Wi(\xi-2)\xi-2)<4\eta_s^2(4Wi^2\xi(2-\xi)-1).
\end{equation}

This criterion determines the nature of transient eigenvalue of the JS model for both monotonic and non-monotonic constitutive curves. The Hopf bifurcation during shear start-up of the JS model is not observed for all shear rates (i.e. $Wi$), but it is only observed for shear rates $(Wi)$ for which the above criteria (Eqs.~\ref{js_hopf_criteria} and \ref{eigen_complex}) is satisfied.
\begin{figure}[ht]
   \subfigure[]{
    \includegraphics[scale=0.28]{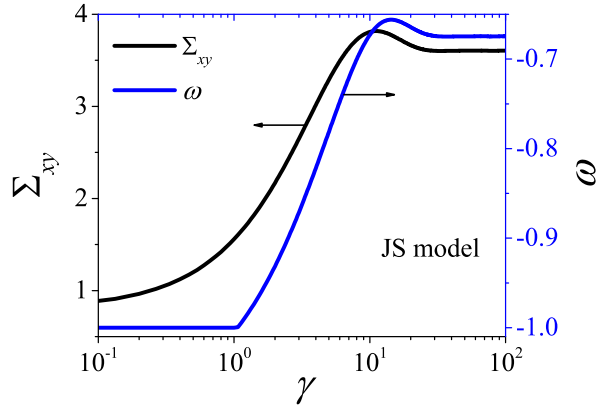}
    \label{js3}
     \quad
  } \subfigure[]{
    \includegraphics[scale=0.28]{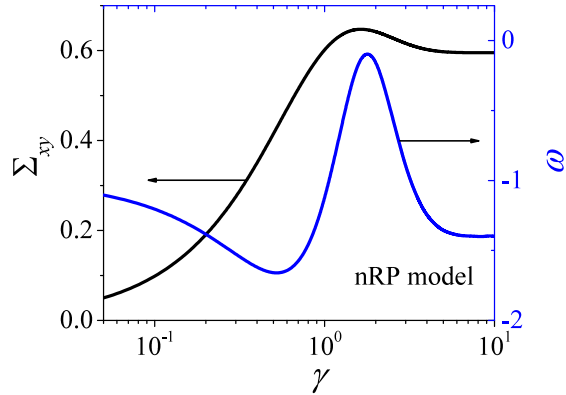}
    \label{nRP3}
  }
  \caption{\small Variation of shear stress (on left y-axis) and eigenvalue (on right y-axis) with shear strain for shear start-up to monotonic region of the flow curve of (a) JS ($Wi=5, ~\eta_s=0.16, ~\xi=0.01$) and (b) nRP ($Wi=3.5,~ \eta_s=10^{-4},~ \beta=0.9$) models. The shear stress shows a decrease with shear strain but no transient unstable eigenvalue is present for both models.}
  \label{fig:jsnRP}
\end{figure}
The demonstration of Hopf bifurcation during $0.68<t<0.8$ runs counter to the earlier work of Moorcroft and Fielding [\onlinecite{moorcroft14}] which explicitly ignored this possibility while deriving the criterion for transient banding during shear start-up. For $0.68<t<0.8$, both the eigenvalues have (identical) positive real parts. This numerical result demonstrates the restrictive nature of the assumption of only one eigenvalue becoming unstable made in the earlier work [\onlinecite{moorcroft14}]. The presence of non-zero imaginary part of the eigenvalue in a system is suggestive of oscillations in the time-evolution of the system. The growth of the oscillations depends upon the nature of the real part. %
However, the oscillations observed in the stress evolution of the JS model can directly be related to the non-zero imaginary part of the eigenvalue. Higher the magnitude of the imaginary part, higher the number of oscillations in the shear or normal stress evolution.

However, the earlier effort towards deriving a criterion for onset of transient banding assumed the absence of Hopf bifurcation, and further considered only one (real) eigenvalue becoming unstable. Our stability results for the JS model, however, clearly demonstrate the occurrence of Hopf bifurcation, and this points to one of the reasons for the criterion proposed in Ref. [\onlinecite{moorcroft14}] for shear start-up being non-universal. As mentioned in the Introduction section, a possibility of Hopf bifurcation in JS model has also been reported by Fielding and Olmsted [\onlinecite{fielding2003kinetics}].

Additionally, the results from the frozen-time analysis reveal no consistent correlation between the duration of shear stress decrease with time and the positivity of the real part of $\omega$  (Fig. \ref{fig:jsnRP}). An interesting observation, also reported for stretching [\onlinecite{adams2011transient}] and non-stretching Rolie-Poly models in literature [\onlinecite{moorcroft14}], is the absence of positive eigenvalue even in the presence of stress overshoot for shear rates that are not in the relatively flat regions of the constitutive curve. Our numerical results for the nRP model, shown in the Fig. \ref{nRP3}, are consistent with these prior results. We plot analogous results for the JS model in Fig. \ref{js3} and find that the eigenvalue is negative during shear start-up flow at lower shear rates in the presence of shear stress overshoot and always positive (temporarily) at higher shear rates during shear start-up flow (not shown here). Hence, our results demonstrate that there is a very little correlation between the shear stress overshoot and positive eigenvalue in the frozen-time stability analysis.
The results presented in Fig.~\ref{fig:jsnRP} show that even in cases where there is no Hopf bifurcation in a given model and if only one eigenvalue becomes unstable in the frozen-time analysis, there is no correlation between the positivity of the eigenvalue and the presence of a negative slope in the stress-strain curve. This is another reason that illustrates the non-universal character of the criterion proposed in the literature [\onlinecite{moorcroft14}].

As discussed in detail in Sec. \ref{section_background}, this disagreement could be attributed to two aspects of the derivation of the criteria in Refs.~[\onlinecite{moorcroft14}] and [\onlinecite{moorcroft2013criteria}]: (i) the assumptions used during the derivation of the criterion restricting the universal applicability of the same, and (ii) the criteria being necessary, but not sufficient conditions for the onset of transient banding. The latter aspect is further demonstrated by the result that despite the presence of a negative slope in the stress-strain curve, the eigenvalues from the frozen-time analysis remained negative (Fig. \ref{fig:jsnRP}). 

An important caution regarding JS model that should be pointed out is the unphysical oscillations in the shear and normal stress evolution at high values of $Wi$ [\onlinecite{larson2013constitutive}]. For the present case of using $\xi=0.01$ and $\eta_s=0.16$, the base state stress starts to show three appreciable overshoots if $Wi>25$, which has not been observed experimentally, and hence been termed as unphysical. The results of Hopf bifurcation in Fig. \ref{fig:JS_hopf} are obtained at $Wi=30$ and is only presented here to show a possibility and one should be wary of such a scenario in other constitutive models. However, the following results of JS model are obtained at lesser value of $Wi$ and the base state stress evolution is not unphysical in that case.

\subsection{Linearized evolution of perturbations}
\begin{figure}
   \subfigure[]{
    \includegraphics[scale=0.24]{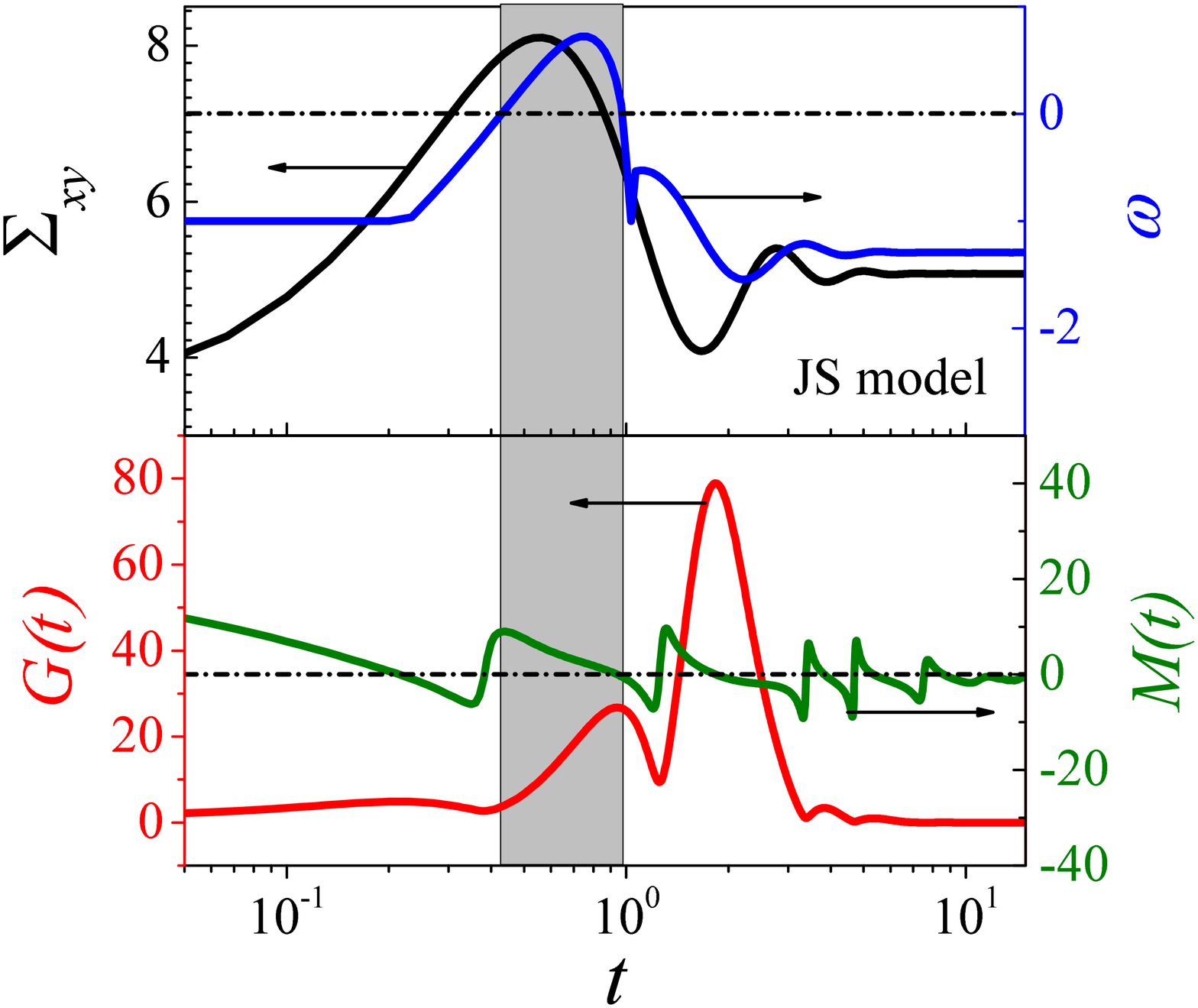}
    \label{lsa_m_js}
  } \subfigure[]{
    \includegraphics[scale=0.24]{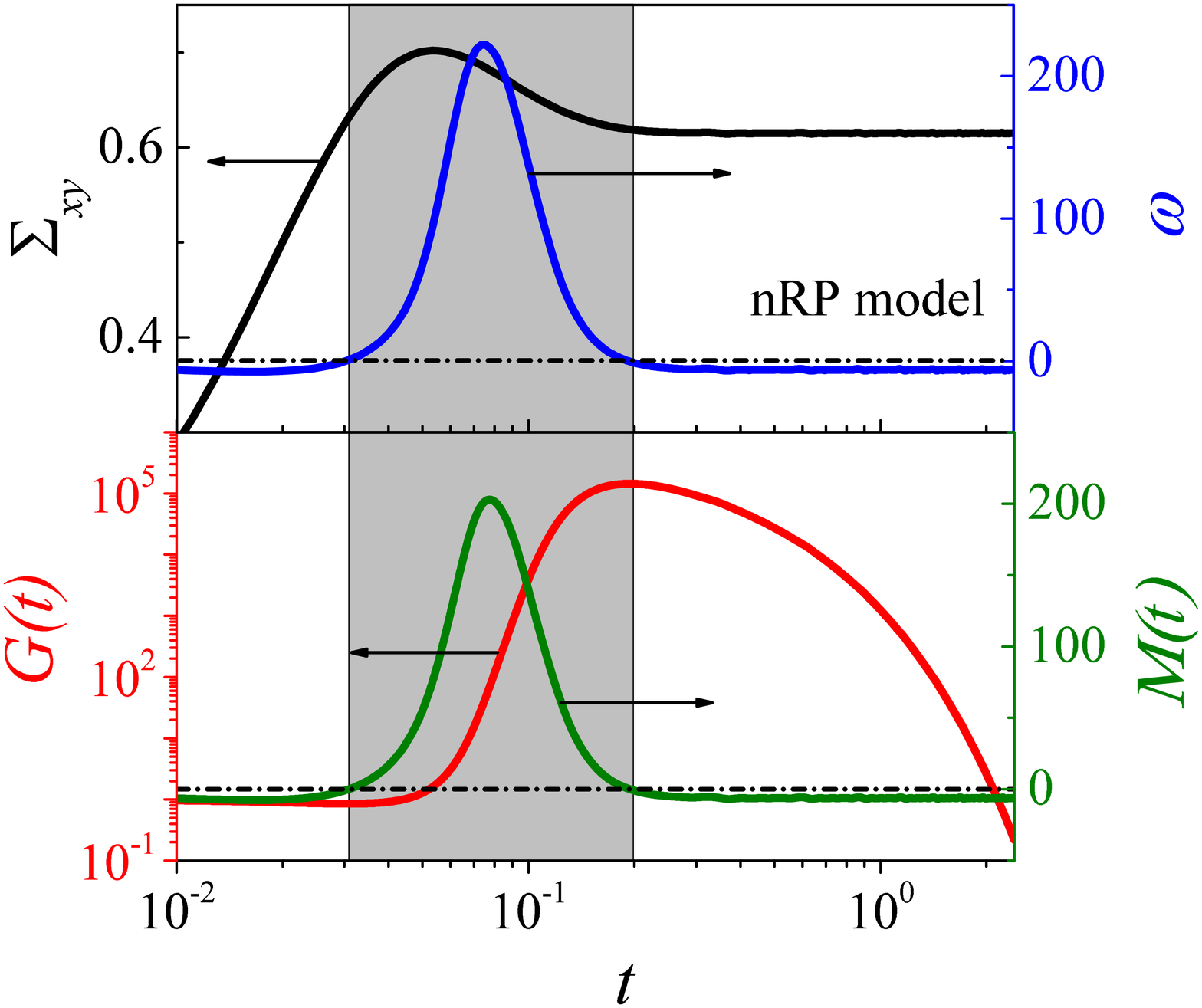}
    \label{lsa_m_nRP}
  }
  \caption{\small Linear stability analysis results for (a) JS model ($Wi=20$ $\eta_s=0.16$) and (b) nRP model ($Wi=30$ $\eta_s=10^{-4}$). The figure (a) and (b) top row shows the evolution of the total shear stress (on left y-axis) and maximum real part of eigenvalue (on right y-axis) with time. In bottom of figure (a) and (b), the evolution of $G(t)$ as a function of time is plotted on left y-axis, which shows the maximum amplification of perturbation at a time $t$, and is independent of initial condition of perturbation. The right y-axis shows evolution of $M(t)$, which represents the actual growth rate of linearized perturbations as $M(t)$ $=$ $\displaystyle\frac{1}{G}\displaystyle\frac{dG}{dt}$. The shaded area in figure (a) and (b) highlights the time duration during which eigenvalue is positive. The black short dash-dot line in top row of figure (a) and (b) shows the $\omega=0$ line  and in the bottom row shows the $M(t)=0$ line.} 
 \label{fig:lsa_m}
\end{figure}

In this subsection, we determine the relationship between a positive transient eigenvalue (obtained from frozen-time analysis), stress overshoot, growth coefficient $G(t)$ and growth rate $M(t)$ as obtained from the linearized evolution of disturbances.
Figure~\ref{fig:lsa_m} shows results from the numerical solution of the linearized equations for the shear start-up flow of the JS and nRP models, as the Giesekus model only reveals negative (stable) eigenvalues in the frozen-time analysis.
Here, figures~\ref{lsa_m_js} and \ref{lsa_m_nRP}, top row, show the evolution of the total shear stress with time and the eigenvalue obtained from the frozen-time analysis.
Figures~\ref{lsa_m_js} and \ref{lsa_m_nRP}, bottom row, illustrate the variation of $G(t)$ and $M(t)$ with time. The shaded region denotes the range of times for which the frozen-time analysis predicts a positive growth rate (i.e. $\omega > 0$).

In Fig. \ref{lsa_m_js}, for the JS model, the shaded region signifies two observations. First, $\omega$ is not positive for the entire duration of the decrease in shear stress with time for the first stress overshoot. Moreover, the subsequent overshoots associated with stress oscillations also do not correspond to positive growth rates from the frozen-time analysis (here, growth rate is the real part of $\omega$). It should be pointed out that Moorcroft and Fielding [\onlinecite{moorcroft14}] do not explicitly mention that the eigenvalue should remain positive for the entire time duration during which the stress decreases with time. However,  the analytical derivation of the criterion (Eq. 54 of Ref. [\onlinecite{moorcroft14}]) implies that the eigenvalue should be positive during the entire interval where the stress decreases with time. More importantly, the rise and decay of $G(t)$ have no correlation to the range of times where the growth rate from the frozen-time analysis is positive. The lack of correlation between $G(t)$ and the the frozen-time eigenvalue demonstrates the limitations of the latter for analyzing the stability of shear start-up, since the numerically obtained linearized evolution shows that the time variation of the base-state stresses seems to influence the  growth coefficient in a significant way. The frozen-time analysis may be expected to be relevant only when the time variation of the base state stresses is much slower compared to the growth rate of the perturbations as obtained from the frozen-time analysis.
A comparison of $M(t)$ with the time variation of $\omega$ suggests that there is a very little correlation between the two.

On the other hand, the results obtained using the nRP model (Figure \ref{lsa_m_nRP}) indicate some correlation between the range of times where the eigenvalue is positive and the times for which $G(t)$ exhibits growth with time (i.e., $M(t)> 0$).
Note, however, that the increase of growth coefficient from the linearized dynamics, the onset of positivity of the growth rate from the frozen-time analysis, and the decrease in shear stress in time do not occur at the same time instant. Interestingly, evolution of $M(t)$ is in close agreement with evolution of the eigenvalue from the frozen-time analysis, in agreement with what was found in Ref.~[\onlinecite{moorcroft14}]. Overall, the frozen-time analysis is not an accurate tool for determining the stability of the transient states considered in this work. Consequently, we also conclude that the use of a frozen-time analysis may not be very accurate in the derivation for the criteria for the onset of transient shear banding. This conclusion is in agreement with the results reported in the literature [\onlinecite{2018PhDT.......101P}]. 

\subsubsection*{Comparison with the inference of Moorcroft and Fielding [\onlinecite{moorcroft14}]}
Here, we highlight the possible connections between the results, discussion and conclusions of Moorcroft and Fielding [\onlinecite{moorcroft14}] in the context of shear start-up and those inferred in the present work. First, Figs.~15 and 16 of Ref. [\onlinecite{moorcroft14}] also presented results from frozen-time analysis and linearized evolution of perturbations for the nRP model. These figures also showed that the frozen time analysis results are not always in agreement with linearized growth and decay of perturbations. This is also in agreement with the results presented in this manuscript using both the frozen-time analysis and the fundamental matrix method for the nRP model. However, we must emphasize again that the Moorcroft and Fielding [\onlinecite{moorcroft14}] concluded that frozen time analysis agreed well with linearized evolution of perturbations in their work, and this conclusion is at variance with our predictions. (We must highlight the fact that Figs.~15 and 16 of Ref. [\onlinecite{moorcroft14}] also do not show complete agreement of results of frozen time analysis and linearized evolution of perturbations and therefore, numerical results of this manuscript and of Ref. [\onlinecite{moorcroft14}] are in agreement, however, the conclusion is not same.) Second, Figs.~15 and 16 of Ref. [\onlinecite{moorcroft14}] also showed that there is no transiently positive eigenvalue at high shear rates for the nRP model. We also show this result for the nRP model, but even at a low shear rates. Interestingly, the same paper by Moorcroft and Fielding also states in its abstract that \textit{``In each case this criterion depends only on the shape of the experimentally measured rheological response function for that protocol, independent of the constitutive properties of the material in question (Therefore our criteria in fact concern all complex fluids and not just the polymeric ones of interest here.)”}. Therefore, despite the possible limitations implied in their Figs. 15 and 16, the authors nevertheless characterized their criterion as being universal. However, our conclusion of the absence of a universal connection between shear stress overshoot and transiently positive eigenvalue using the JS and nRP models is at variance with the original papers of Moorcroft and Fielding [\onlinecite{moorcroft14,moorcroft2013criteria}], but is in broad agreement with the caveat subsequently expressed  by Fielding [\onlinecite{fielding2016triggers}] regarding  generality of the criterion for shear start-up.

\subsection{Role of flatness of constitutive curve}

\begin{figure}
     \subfigure[]{
    \includegraphics[scale=0.25]{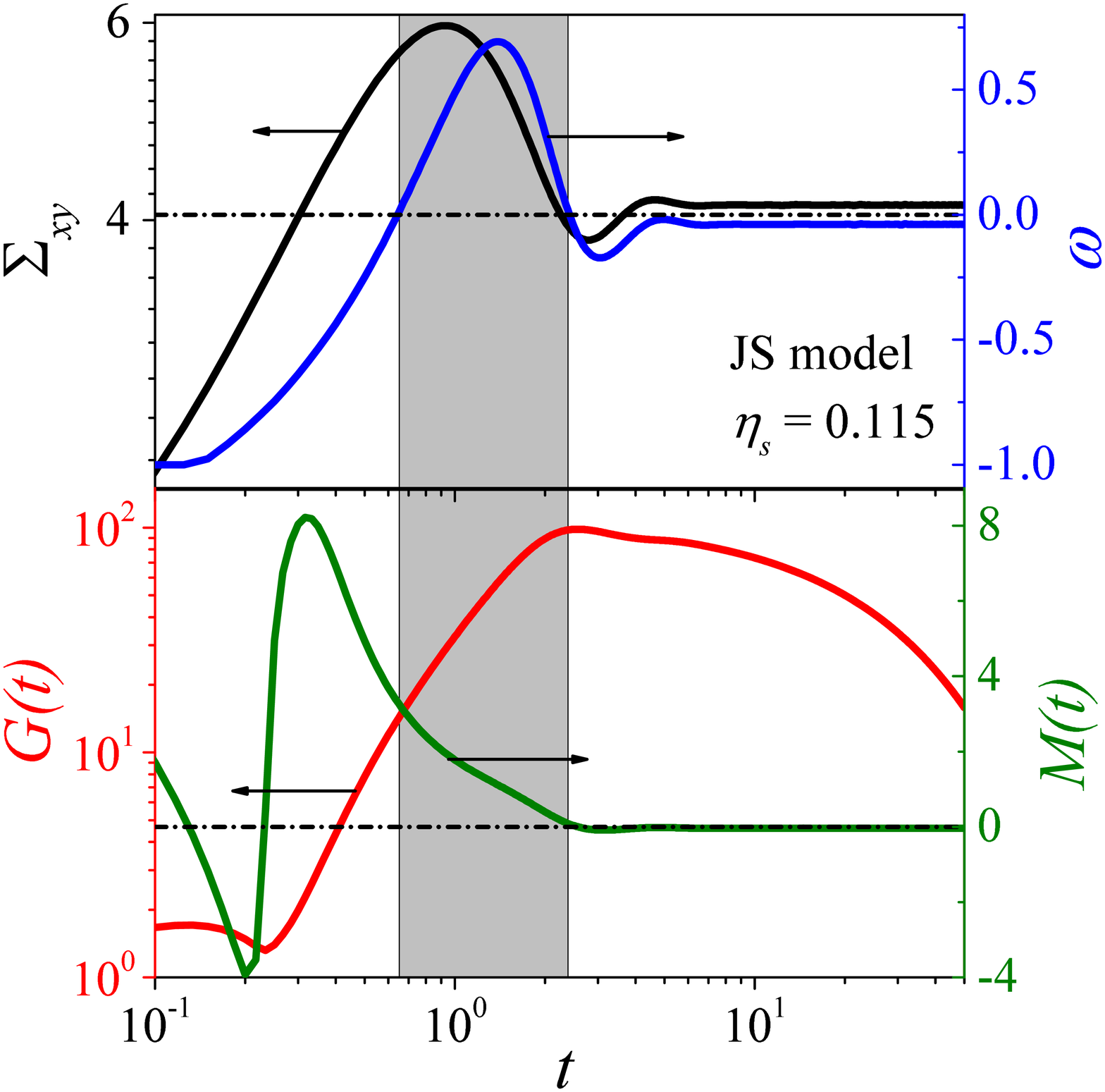}
   
    \label{fig:js_flat_0.115}
    } \subfigure[]{
    \includegraphics[scale=0.25]{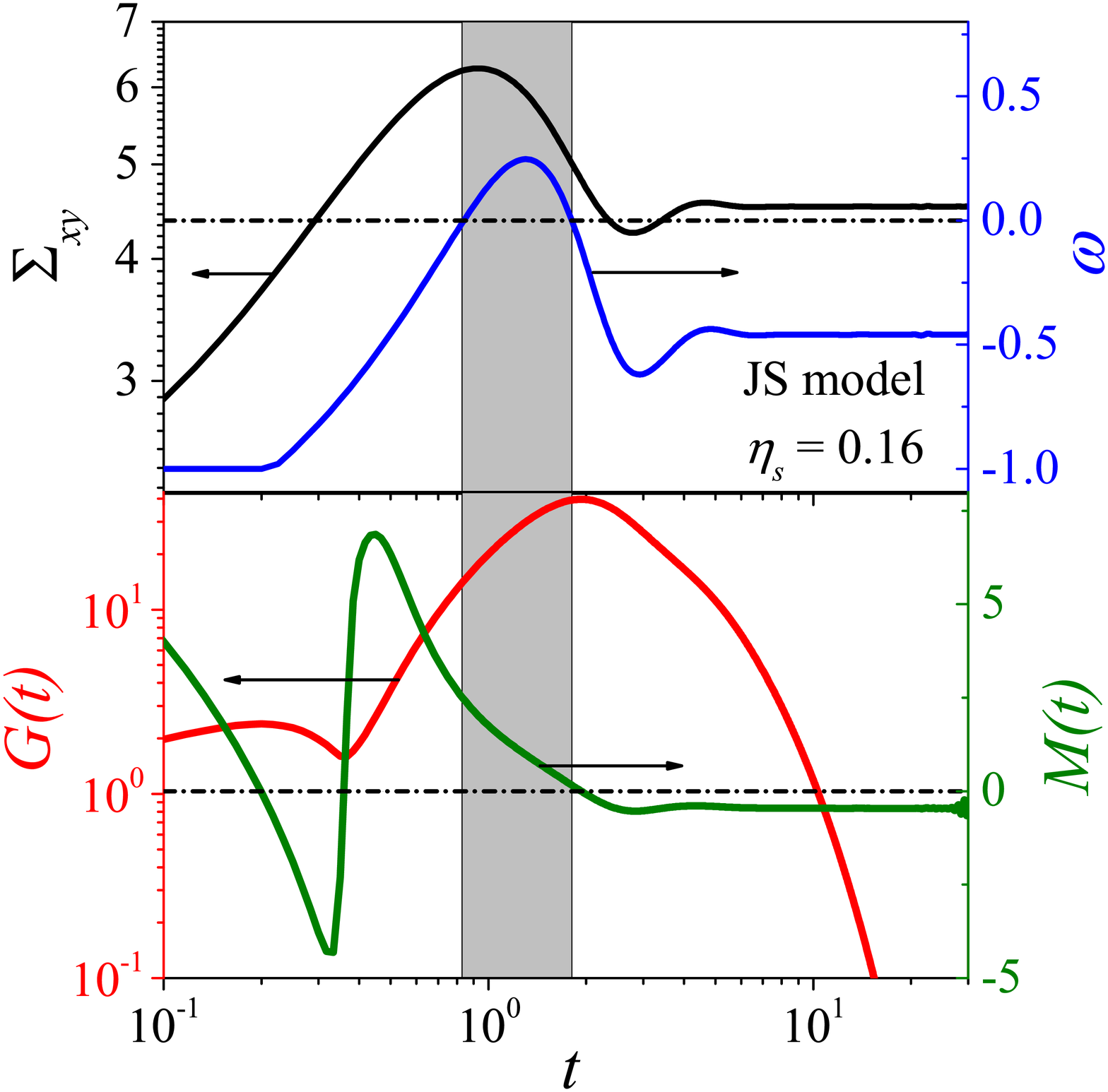}
   
    \label{fig:js_flat_0.16}
    }
     \caption{\small Linear stability analysis results for JS model at $Wi=12$ for (a) $\eta_s=0.115$ and (b) $\eta_s =0.16$. The figure (a) and (b) top row shows the evolution of total shear stress (on left y-axis) and maximum real part of eigenvalue (on right y-axis) time. In bottom of figure (a) and (b), the evolution of $G(t)$ as a function of time is plotted on the left y-axis, which shows the maximum amplification of perturbation at a time $t$ and is independent of initial condition of perturbation. The right y-axis shows evolution of $M(t)$, which represents the actual growth rate of linearized perturbations as $M(t)$ $=$ $\displaystyle\frac{1}{G}\displaystyle\frac{dG}{dt}$. The shaded area in figure (a) and (b) highlights the time duration during which eigenvalue is positive. The black short dash-dot line in top row of figure (a) and (b) represents the $\omega=0$ while in the bottom row represents that corresponding to $M(t)=0$.} 
    \label{fig:js_flat}
\end{figure}

Adams and Olmsted [\onlinecite{adams2009nonmonotonic}] and Moorcroft and Fielding [\onlinecite{moorcroft14}] shared a common observation that the shear start-up flow is transiently unstable if the shear rate lies in the nearly flat region of the constitutive curve. The authors argued that the flattest region is very close to the inception of the steady-state banding instability of a non-monotonic constitutive curve, and therefore, the magnitude of the unstable (transient) eigenvalue was also observed to increase with increase in flatness of the constitutive curve. Hence, according to frozen-time analysis, the system was concluded to be transiently unstable for shear start-up flow with shear rate in the near-flat part of the constitutive curve. Interestingly, the flatness of the constitutive curve increases by decreasing the value of $\eta_s$ and all these models become pathological at $\eta_s=0$ as shown in Fig. \ref{n_0_results}. The decrease in $\eta_s$ also increases width of the flat portion of the constitutive curve. We explore this finding further using JS, nRP and Giesekus models below.
In Figs. \ref{fig:js_flat_0.115} and \ref{fig:js_flat_0.16}, we plot the base-state shear stress, transient eigenvalue, growth coefficient, $G(t)$ and growth rate, $M(t)$ with time for shear start-up flow of JS model with $\eta_{s}=0.115$ and $\eta_{s}=0.16$, respectively. The value of $\left(\displaystyle \frac{d\Sigma_{xy}}{dWi}\right)$ is 0.0045 and 0.055, respectively for the two cases. Apart from the aforementioned observation of no direct correlation between the duration of positive growth rate from the frozen-time analysis, the negative slope in the stress-strain curve, the growth and decay of linearized perturbations growth coefficient and evolution of $M(t)$, we find that with a decrease of $\eta_s$ from 0.16 to 0.115, the maximum transient growth rate from the frozen-time analysis increases from 0.1 to 0.6 and the maximum value of the $G(t)$ increases from 40 to 100. The transient maximum value of $M(t)$ also increases from 5 to 8.

\begin{figure}
     \subfigure[]{
    \includegraphics[scale=0.2]{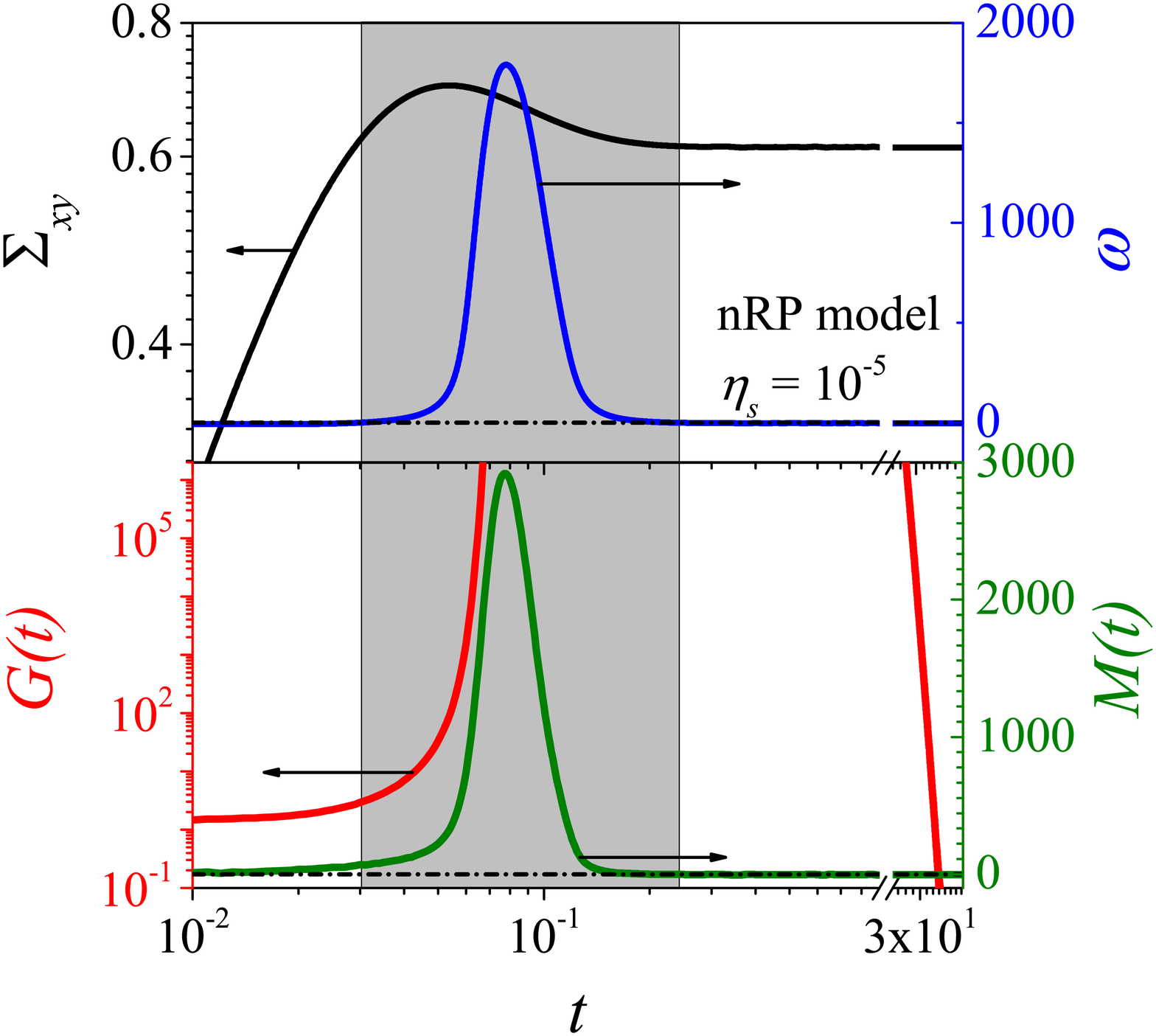}
   
    \label{fig:nrp_flat_1e-5}
    } \subfigure[]{
    \includegraphics[scale=0.2]{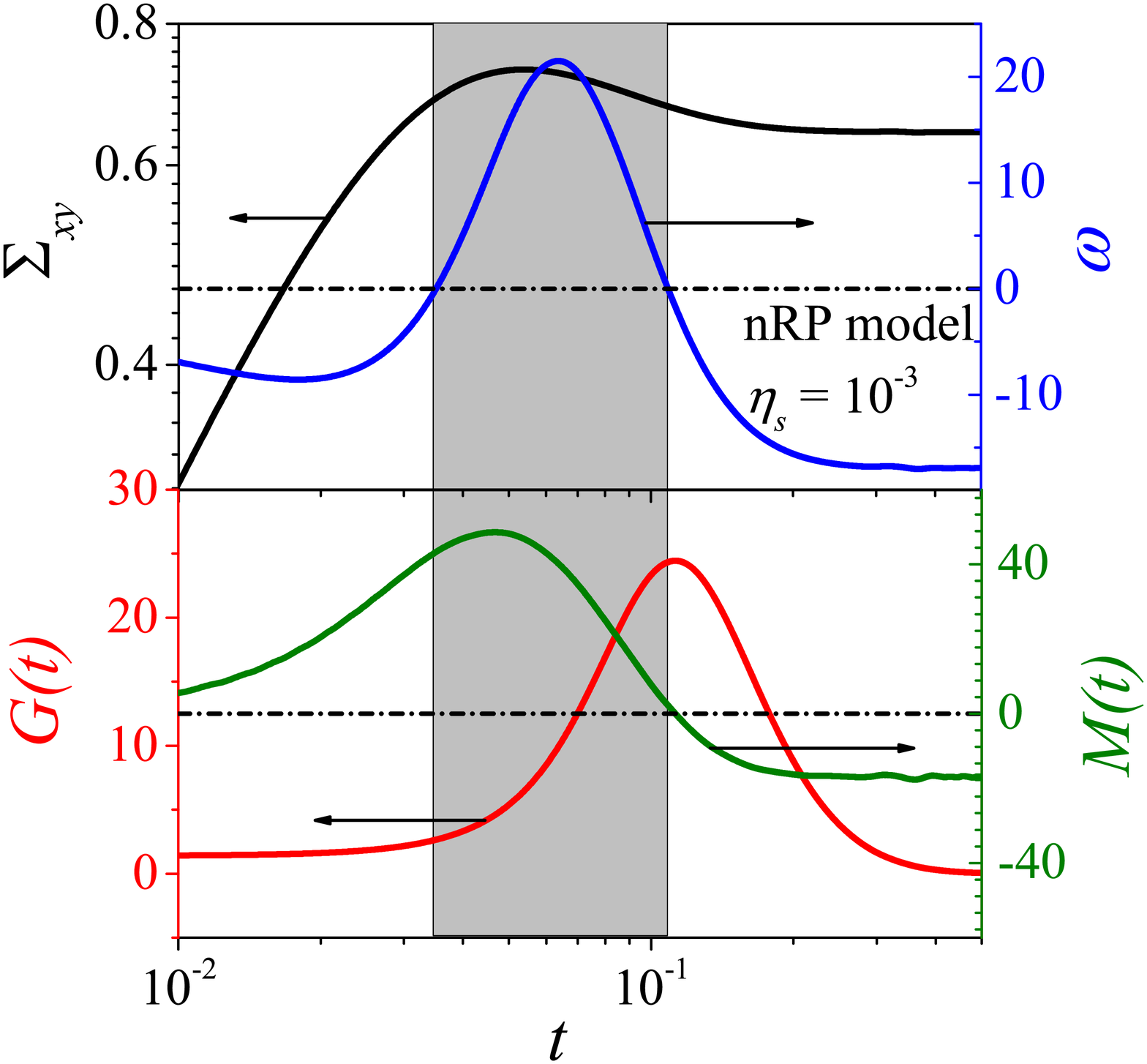}
   
    \label{fig:nrp_flat_1e-3}
    }
     \subfigure[]{
    \includegraphics[scale=0.22]{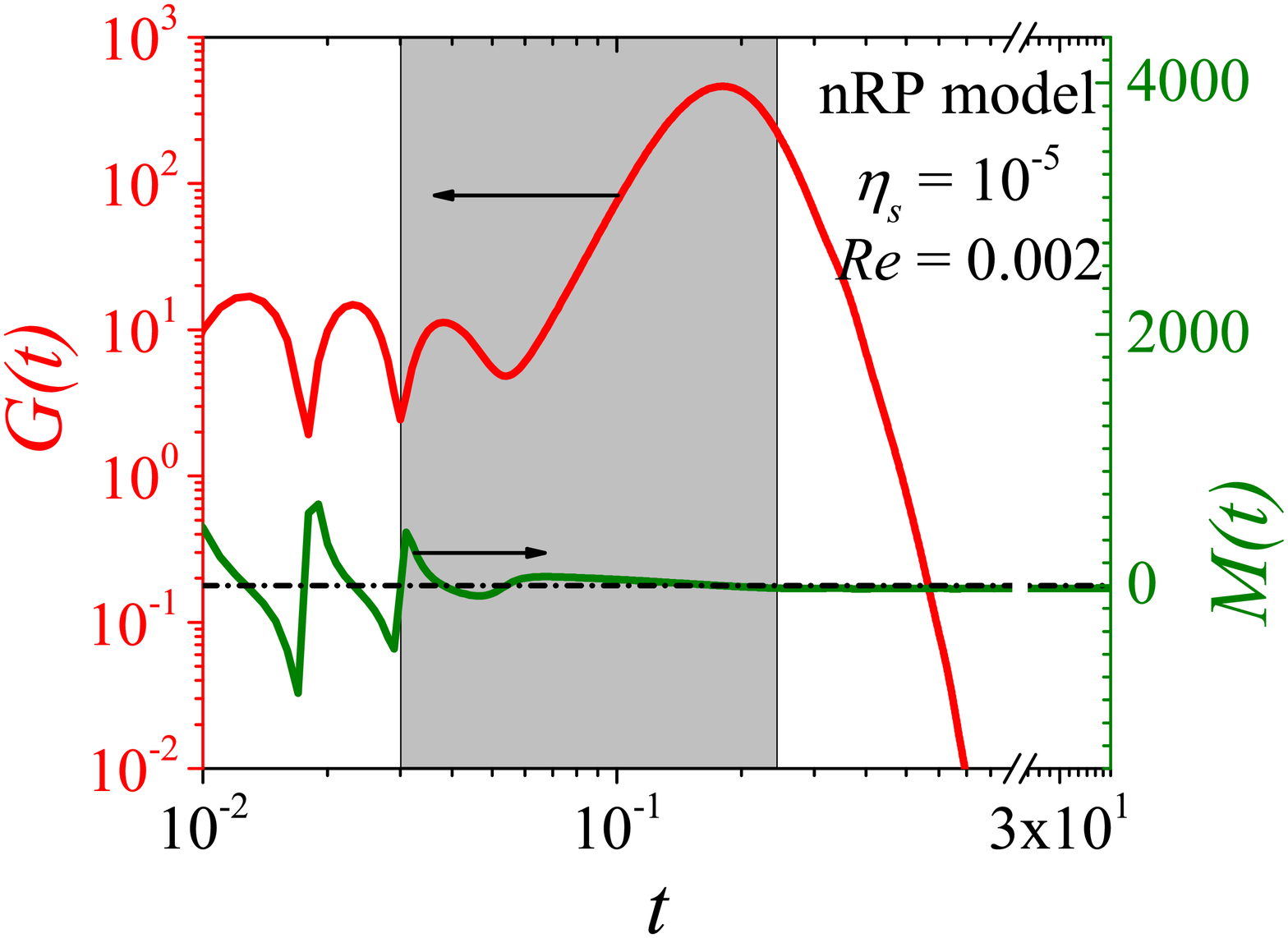}
   
    \label{fig:nrp_flat_1e-5_inertia}
    }
      \subfigure[]{
    \includegraphics[scale=0.22]{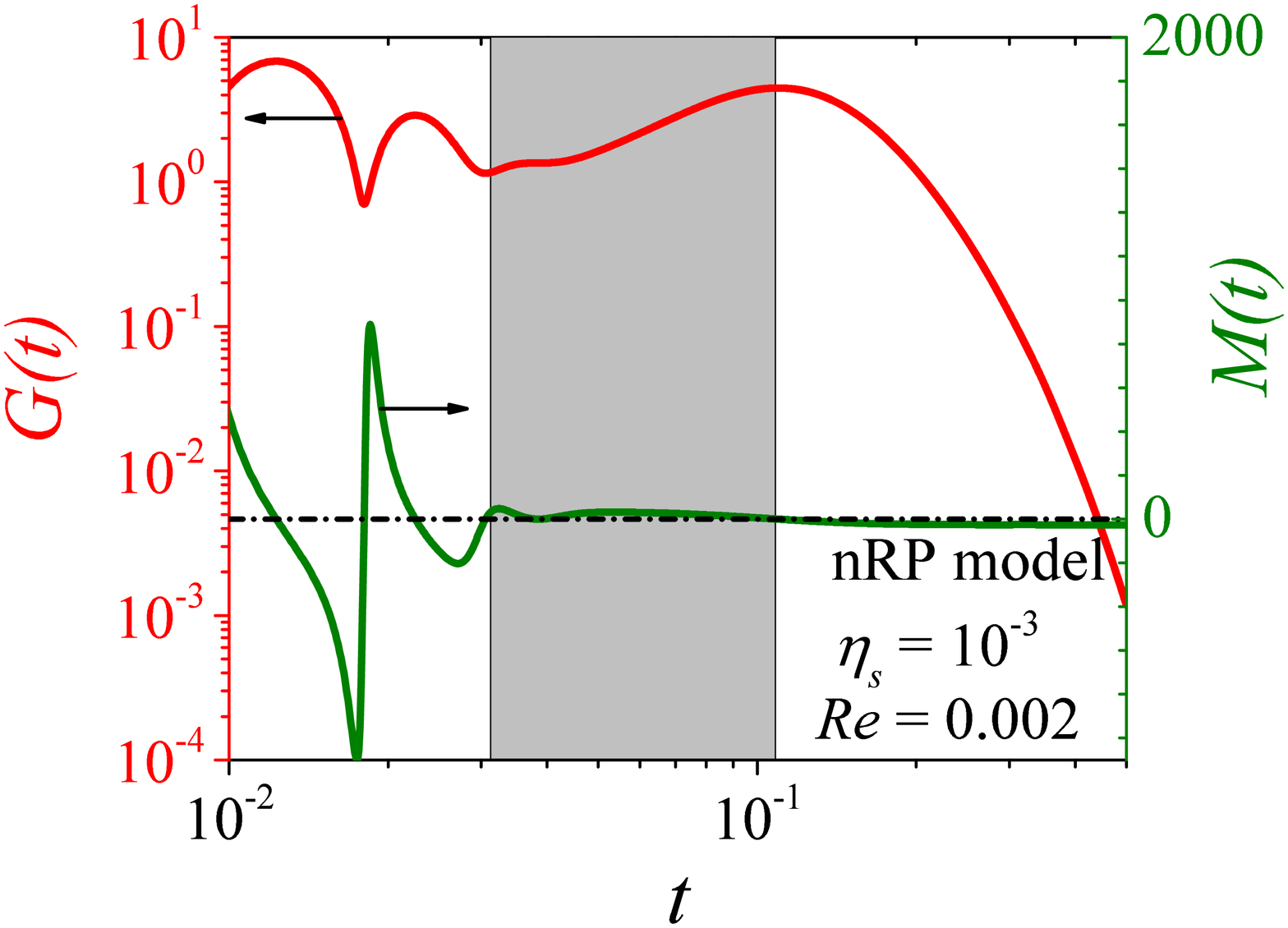}
   
    \label{fig:nrp_flat_1e-3_inertia}
    }
     \caption{\small Linear stability analysis results for nRP model at $Wi=30$ for (a) $\eta_s=10^{-5}$ and (b) $\eta_s =10^{-3}$. The figure (a) and (b) top row shows the total shear stress (on left y-axis) and maximum real part of eigenvalue (on right y-axis) evolution with time. In bottom of figure (a) and (b), the evolution of $G(t)$ as a function of time is plotted on the left y-axis which shows the maximum amplification of perturbation at a time $t$ and is independent of initial condition of perturbation. The right y-axis shows evolution of $M(t)$, which represents the actual growth rate of linearized perturbations as $M(t)$ $=$ $\displaystyle\frac{1}{G}\displaystyle\frac{dG}{dt}$. The shaded area in figure (a) and (b) highlights the time duration during which eigenvalue is positive. The black short dash-dot line in top row of figure (a) and (b) represents $\omega=0$ line while in the bottom row represents $M(t)=0$. The figure (c) and (d) shows the evolution of $G(t)$ and $M(t)$ in presence of inertia, $Re=2\times 10^{-3}$ and at $n=1$ corresponding to the parameters of the creeping flow results in figure (a) and (b), respectively. In figure (a), bottom row, the value of $G(t)$ diverges transiently to $\sim$ $10^{30}$ and not shown on the figure due to limitation of scale. This divergence shows the limitations of a linearized study.}
    \label{fig:nrp_flat}
\end{figure}
We also plot results for the nRP model in Fig. \ref{fig:nrp_flat_1e-5} and \ref{fig:nrp_flat_1e-3} for shear start-up flow for $\eta_{s}=10^{-5}$ and $\eta_{s}=10^{-3}$, respectively. The value of $\left(\displaystyle \frac{d\Sigma_{xy}}{dWi}\right)$ is $4\times10^{-5}$ and $1\times10^{-3}$, respectively for the two cases. Again, the nRP model shows some correlation between duration of positive growth rate (from the frozen-time analysis), negative slope of stress-strain curve and $M(t)$, but no correlation between negative slope of stress-strain curve and growth and decay of $G(t)$. The maximum transient value of growth rate from the frozen-time analysis, $\omega$ increases from 20 to 2000, maximum value of $G(t)$ diverges from 25 to $10^{30}$, and maximum value of $M(t)$ increases from 40 to 3000, on decreasing $\eta_{s}$ from $10^{-3}$ to $10^{-5}$. (Figures \ref{fig:nrp_flat_1e-5_inertia} and \ref{fig:nrp_flat_1e-3_inertia} also shows results for linearized perturbations growth coefficient evolution for nRP model in the presence of inertia which are discussed in Sec. \ref{inertia_effect}.)

\begin{figure}
     \subfigure[]{
    \includegraphics[scale=0.25]{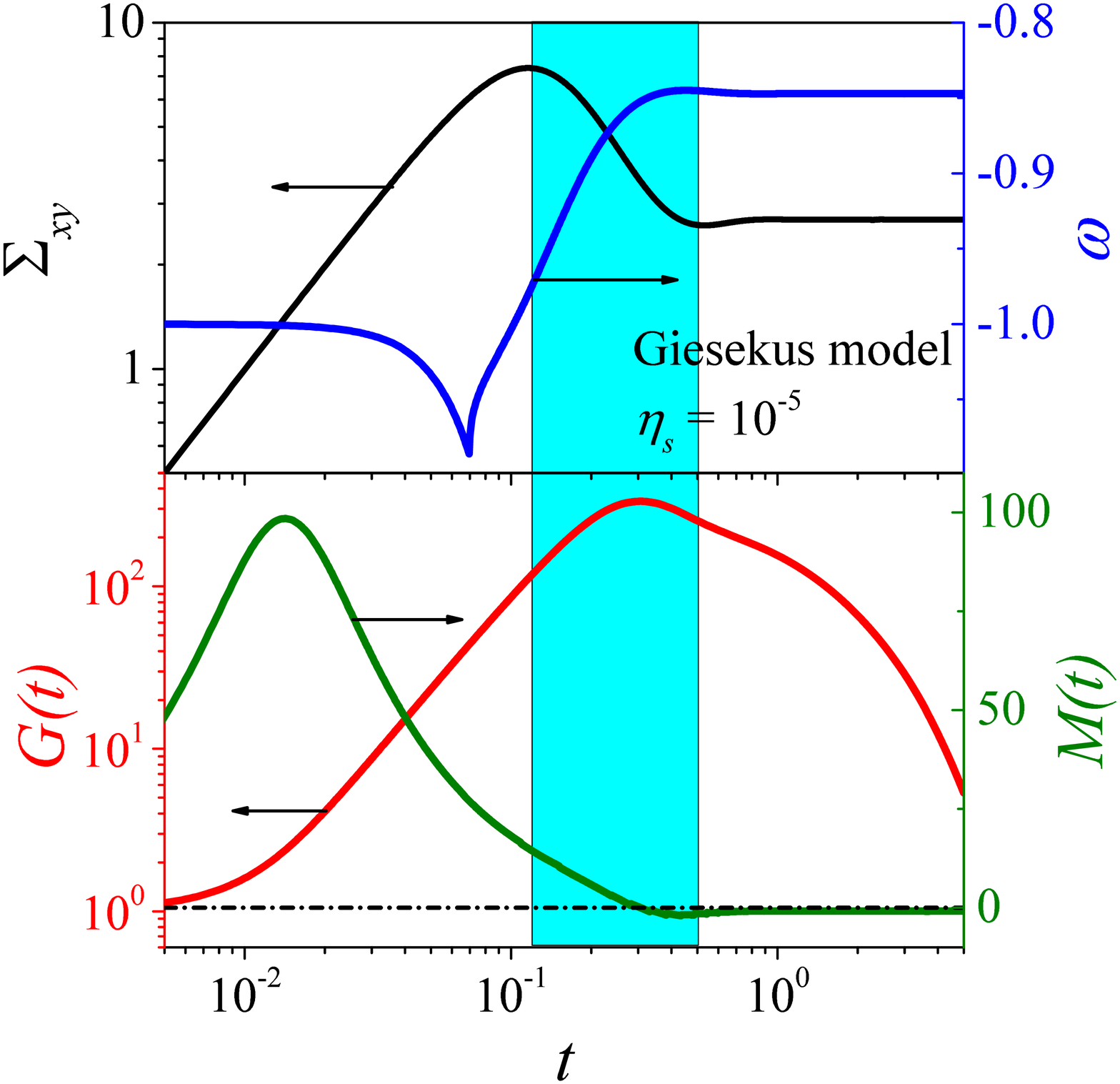}
   
    \label{fig:g_flat_1e-5}
    } \subfigure[]{
    \includegraphics[scale=0.25]{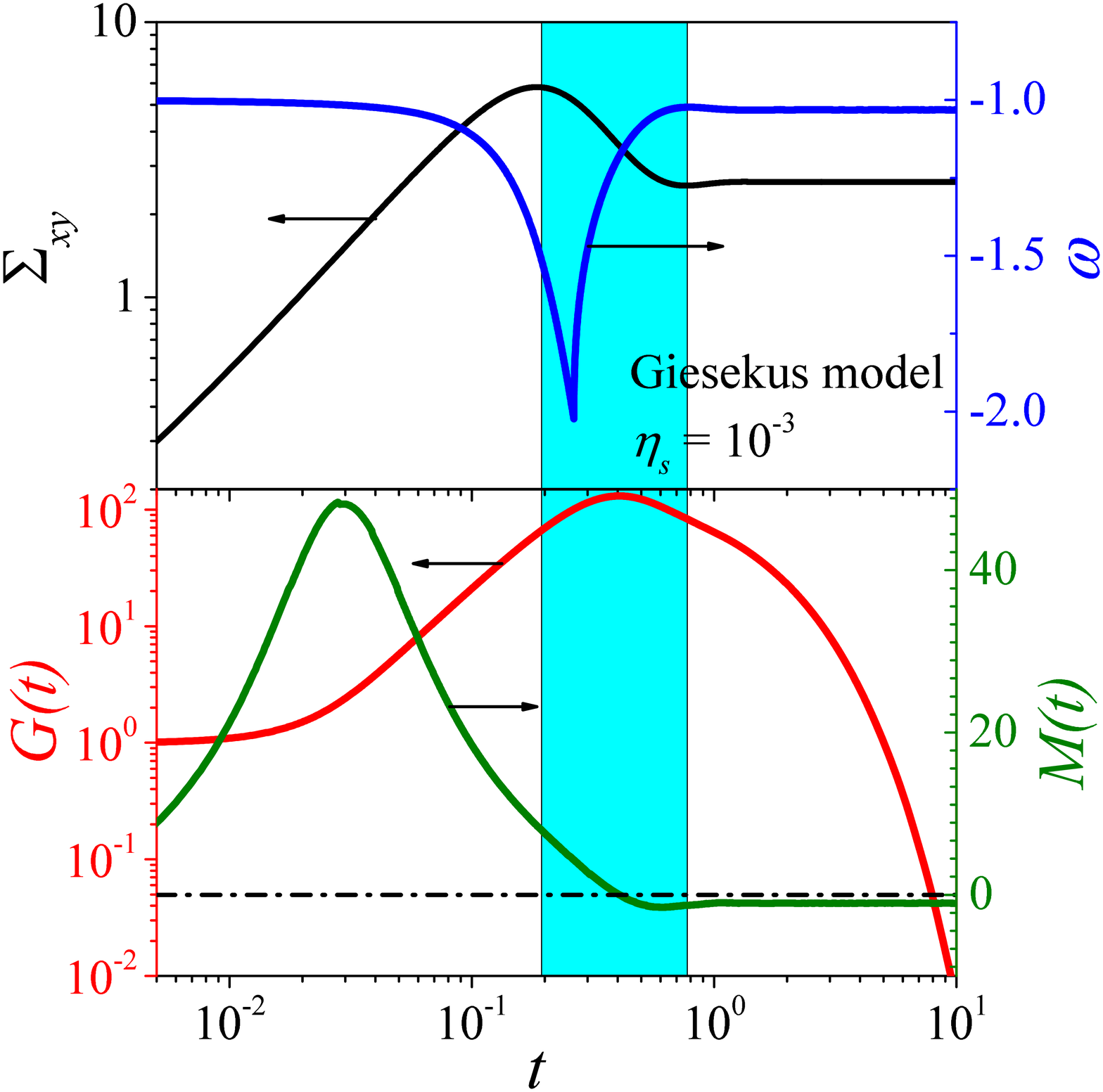}
   
    \label{fig:g_flat_1e-3}
    }
     \caption{\small Linear stability analysis results for Giesekus model with $\alpha=0.1$ at (a) $Wi=100$ for $\eta_s=10^{-5}$ and at (b) $Wi=50$ for $\eta_s =10^{-3}$. The figure (a) and (b) top row shows the total shear stress (on left y-axis) and maximum real part of eigenvalue (on right y-axis) evolution with time. In bottom of figure (a) and (b), the evolution of $G(t)$ as a function of time is plotted on the left y-axis, which shows the maximum amplification of perturbation at a time $t$ and is independent of initial condition of perturbation. The right y-axis shows evolution of $M(t)$, which represents the actual growth rate of linearized perturbations as $M(t)=\displaystyle\frac{1}{G}\displaystyle\frac{dG}{dt}$. The shaded area in figure (a) and (b) highlights the time duration during which stress is a decreasing function of time. The black short dash-dot line in bottom row of figure (a) and (b) represents the $M(t)=0$ line.}
    \label{fig:g_flat}
\end{figure}

The results obtained for the Giesekus model (Figs. \ref{fig:g_flat_1e-5} and \ref{fig:g_flat_1e-3}), interestingly, show that the eigenvalues are always negative even in the presence of a stress overshoot and in the near-flat region of the constitutive curve, as also noted in Ref. [\onlinecite{moorcroft14}]. Therefore, the effect of flatness of the constitutive curve on linearized evolution of perturbations can be directly identified. We study shear start-up flow for Giesekus model for $\eta_{s}=10^{-5}$, $Wi=100$ and $\eta_{s}=10^{-3}$, $Wi=50$. (We choose $Wi$ such that shear rate lies in the midst of the least steeper portion of the constitutive curve.) We find that growth and decay of $G(t)$ with time even with transiently negative eigenvalues and its maximum value increases approximately from 130 to 300. Interestingly, the evolution of $G(t)$ and $M(t)$ show no appreciable difference with change in $\eta_s$ from $10^{-3}$ to $10^{-5}$ for the same values of $Wi$, i.e., shear start-up flow at $Wi=50$ and $Wi=100$. More importantly, the decrease in stress with time after overshoot does not trigger the growth of perturbation which is clearly visible by the early time positive value and the peak value of $M(t)$ as shown in Fig. \ref{fig:g_flat}.

\begin{figure}
   \subfigure[]{
    \includegraphics[scale=0.25]{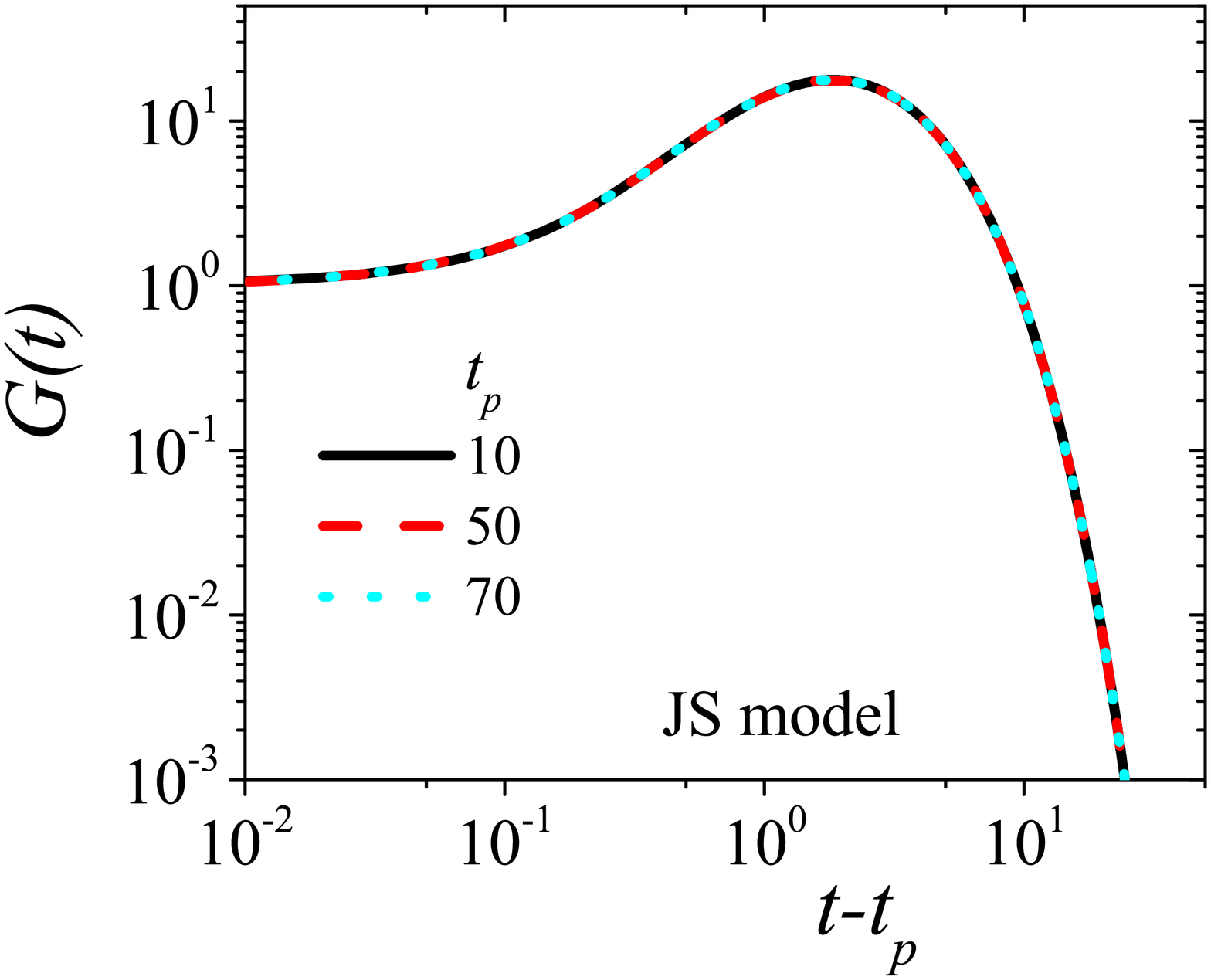}
    \label{js_tp}
    }
    
    \subfigure[]{
    \includegraphics[scale=0.25]{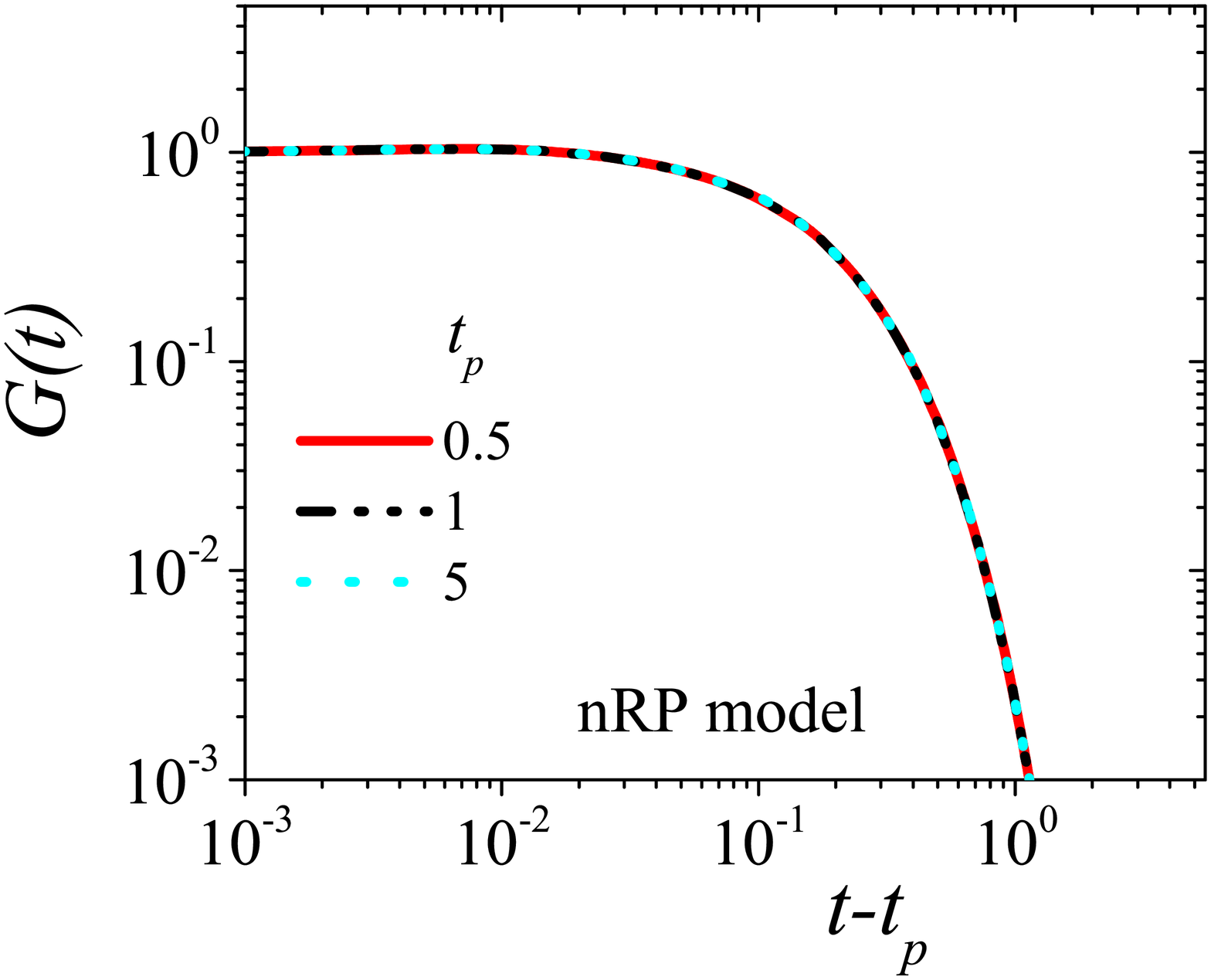}
    \label{nrp_tp}
    }
    
    \subfigure[]{
    \includegraphics[scale=0.25]{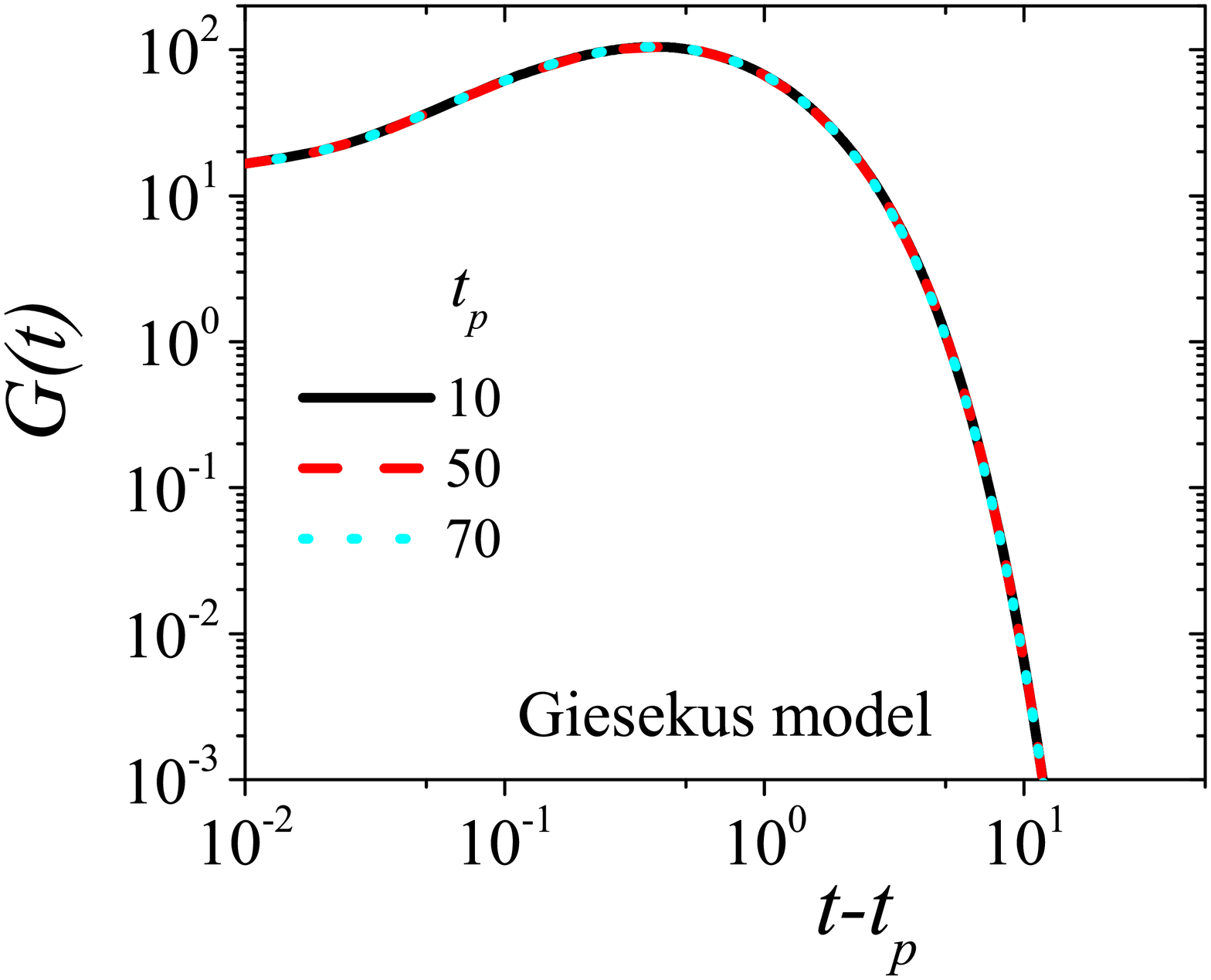}
    \label{giesekus_tp}
    }
    \caption{Effect of time of seeding of perturbation, $t_p$ if $t_p>t_{steady-state}$ for JS model ($Wi=12$, $\eta_s=0.16$) (a), nRP model ($Wi=30$, $\eta_s=10^{-4}$) (b), and Giesekus model ($Wi=50$, $\eta_s=10^{-3}$) (c). The JS and Giesekus models show transient growth even if base state is steady. The nRP model shows a gradual decay of perturbation without any transient growth with base state at steady state.}
    \label{fig:tp_effect}
\end{figure}

We also study the effect of the time of seeding of perturbation, $t_p$ as shown in Fig. \ref{fig:tp_effect} (For all other results of linearized evolution of perturbations presented in this manuscript, $t_p=0$). For $t<t_p$, there is no imposed perturbation during base state evolution. At $t=t_p$, the perturbations are imposed to base state evolution. Figure \ref{fig:tp_effect} shows results for the case of $t_p>t_{steady-state}$, where $t_{steady-state}$ is time at which base state evolution has attained steady state (as determined by the saturation of the stress components with time in the base state). Results from the JS and Giesekus models show a transient growth of perturbations, even if the base state is steady (Fig. \ref{js_tp} and \ref{giesekus_tp}). On the other hand, results from the nRP model show a monotonic decay of perturbations without any transient growth as shown in Fig. \ref{nrp_tp}. If $t_p<t_{steady-state}$, then the transient growth of perturbations is several orders of magnitude higher than the case of $t_p>t_{steady-state}$ for nRP model as shown in Figs. \ref{lsa_m_nRP} and \ref{fig:nrp_flat} for the case of $t_p=0$. For the JS and Giesekus models, the order of magnitude of the growth remains almost similar for cases of $t_p<t_{steady-state}$ and $t_p>t_{steady-state}$. This result is also similar to results obtained for steady plane Couette flow of an Oldroyd-B model [\onlinecite{jovanovic2010transient}]. The growth, in $G(t)$, for perturbations about steady plane Couette flow is referred to as "transient growth" in the field of hydrodynamic stability, which denotes growth in perturbations even though as per the eigenvalue analysis, the flow is stable at large times. 

\begin{figure}
    
    \includegraphics[scale=0.3]{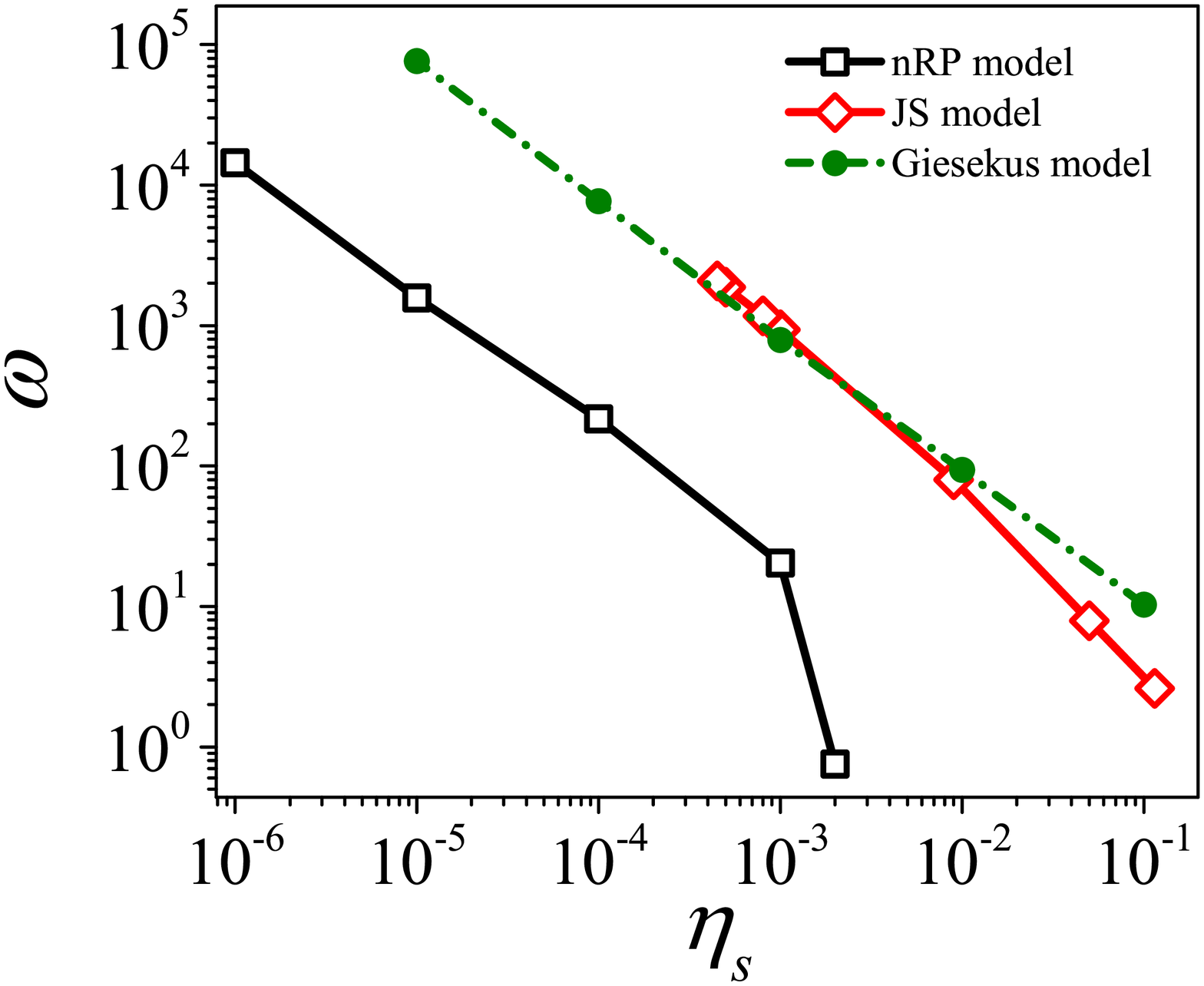}
     \caption{Variation of the eigenvalue as a function of ratio of solvent viscosity to zero shear viscosity of the solution for nRP model, JS model and Giesekus model. For nRP model, $\omega$ is the transient eigenvalue (represented by black open symbol and solid line) at $Wi=30$ and $t=0.067$ for all the values of $\eta_s$. For JS model, $\omega$ is the transient eigenvalue (represented by red open symbol and solid line) at $Wi=350$ and $t=0.062$. For Giesekus model, $\omega$ is modulus of the eigenvalue to represent the decay rate using green closed symbols and dashed lines for all values of $\eta_s$ at $Wi=100$ and $t=0.1$.}
       \label{fig:eigenvalue_diverge}
\end{figure}

In this subsection, we first discuss how the results from the frozen-time analysis are linked to the flatness of the constitutive curve. The JS and nRP models show an increase in the maximum transient eigenvalue with increasing flatness of the constitutive curve. The increase in eigenvalue for nRP model is much larger compared to the JS model even though shear start-up flow of JS model is also in the near-flat region of the constitutive curve. The flatness of the JS constitutive curve is near its maximum extent or $\left(\displaystyle \frac{d\Sigma_{xy}}{dWi}\right)_{min}$ because to increase the flatness, $\eta_s$ has to be decreased below $1/9$ whereupon the constitutive curve becomes non-monotonic as shown in Fig. \ref{ds_dW}. Therefore, the results of JS model shown in Fig. \ref{fig:eigenvalue_diverge} are obtained using $Wi=350$. It must be noted that the high value of $Wi$ yields unphysical oscillations in the stress evolution. The value of $Wi=350$ is only chosen to show the trend of $\omega$ with $\eta_s$, which is not possible at low value of $Wi$ but the trend is same up to $\eta_s>1/9$ for lower values of $Wi$ also. This is because of the limitations associated with lowering the value of $\eta_s$ that leads to shear rate in the lower monotonic region of the constitutive curve, which shows a transiently positive eigenvalue in presence of the stress overshoot. For the case of Giesekus model, the transient eigenvalue is always negative and there is no effect of flatness of the constitutive curve on the growth rate from the frozen-time analysis. There is an effect of flatness of constitutive curve on the decay rate which will be discussed later. 
\begin{figure}
    \includegraphics[scale=0.3]{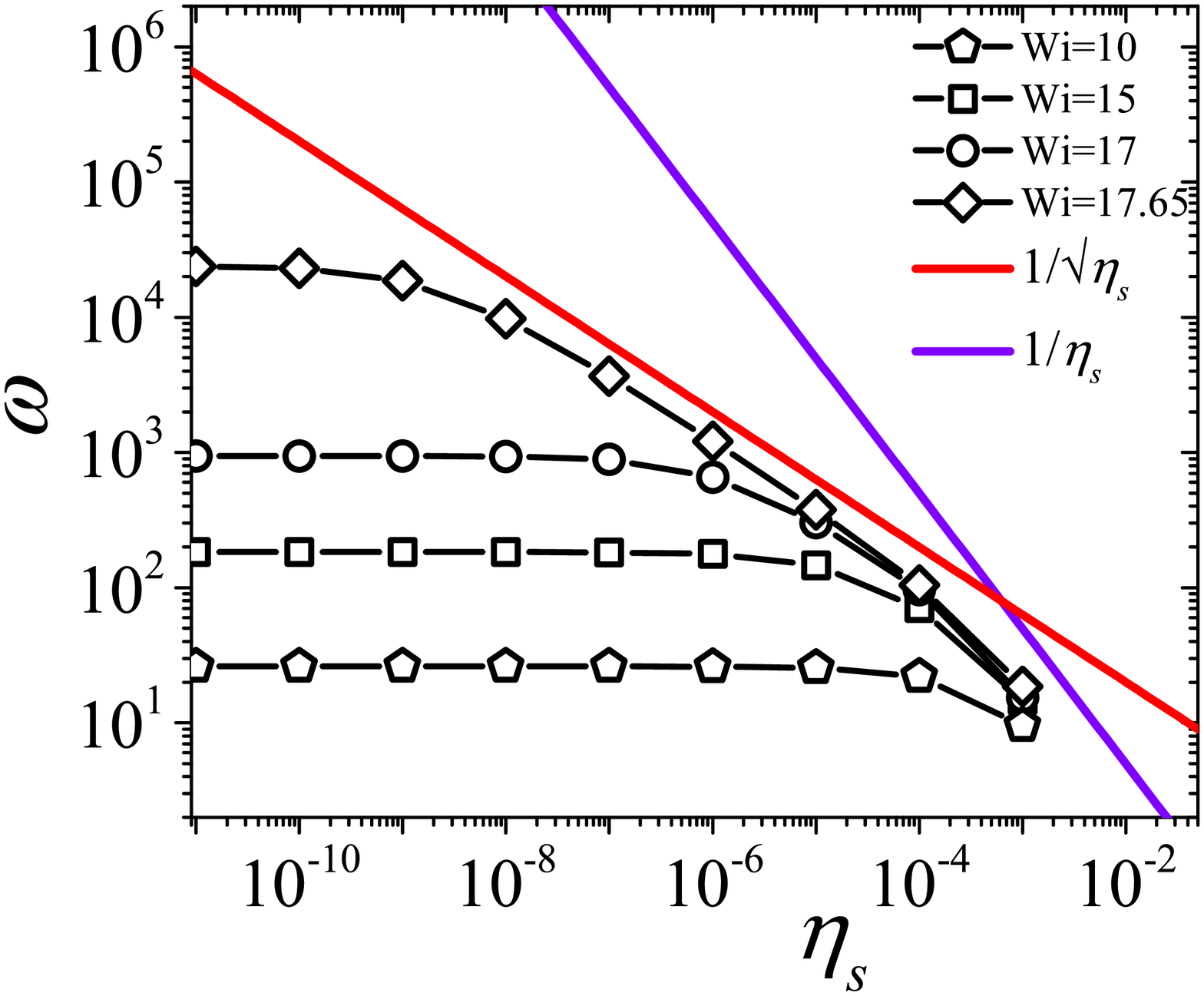}
    \caption{Variation of transiently maximum eigenvalue as a function of $\eta_s$ for the case of nRP model using $\beta=1$ at $Wi=10$ (at $t=0.21$), $Wi=15$ (at $t=0.15$ ), $Wi=17$ (at $t=0.13$ ), and $Wi=17.65$ (at $t=0.13$). This result shows that in absence of strong flows, $\omega$ plateaus to a constant value on decreasing $\eta_s$. The scaling followed before plateauing is $\omega\sim\eta_s^{-1/2}$ if $Wi<18$.}
    \label{fig:eigenvalue_diverge_nRP}
\end{figure}

\begin{figure}
    \subfigure[]{
    \includegraphics[scale=0.28]{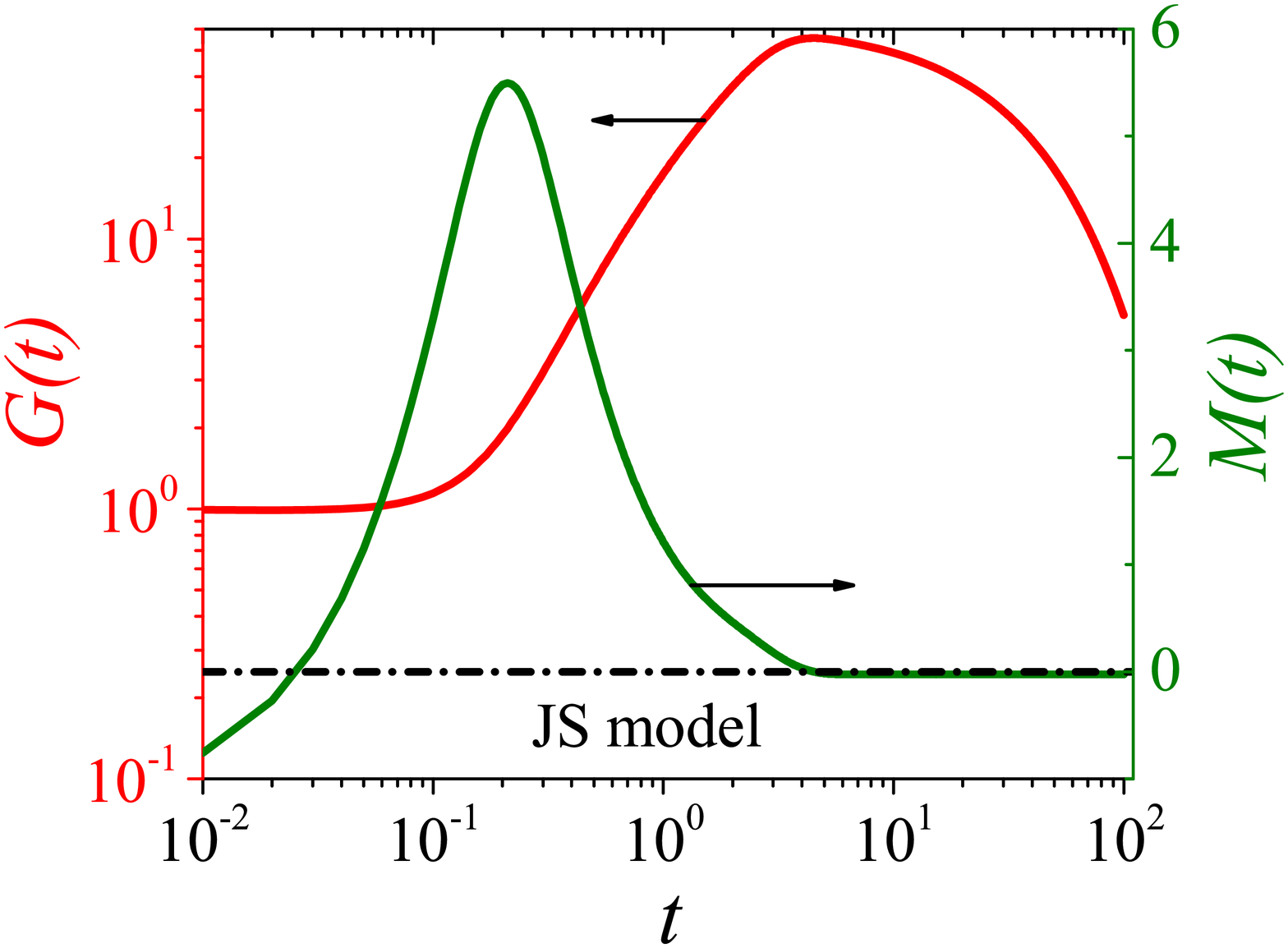}
    
    \label{fig:js_n_0_g_m}
    }
    
    \subfigure[]{
    \includegraphics[scale=0.28]{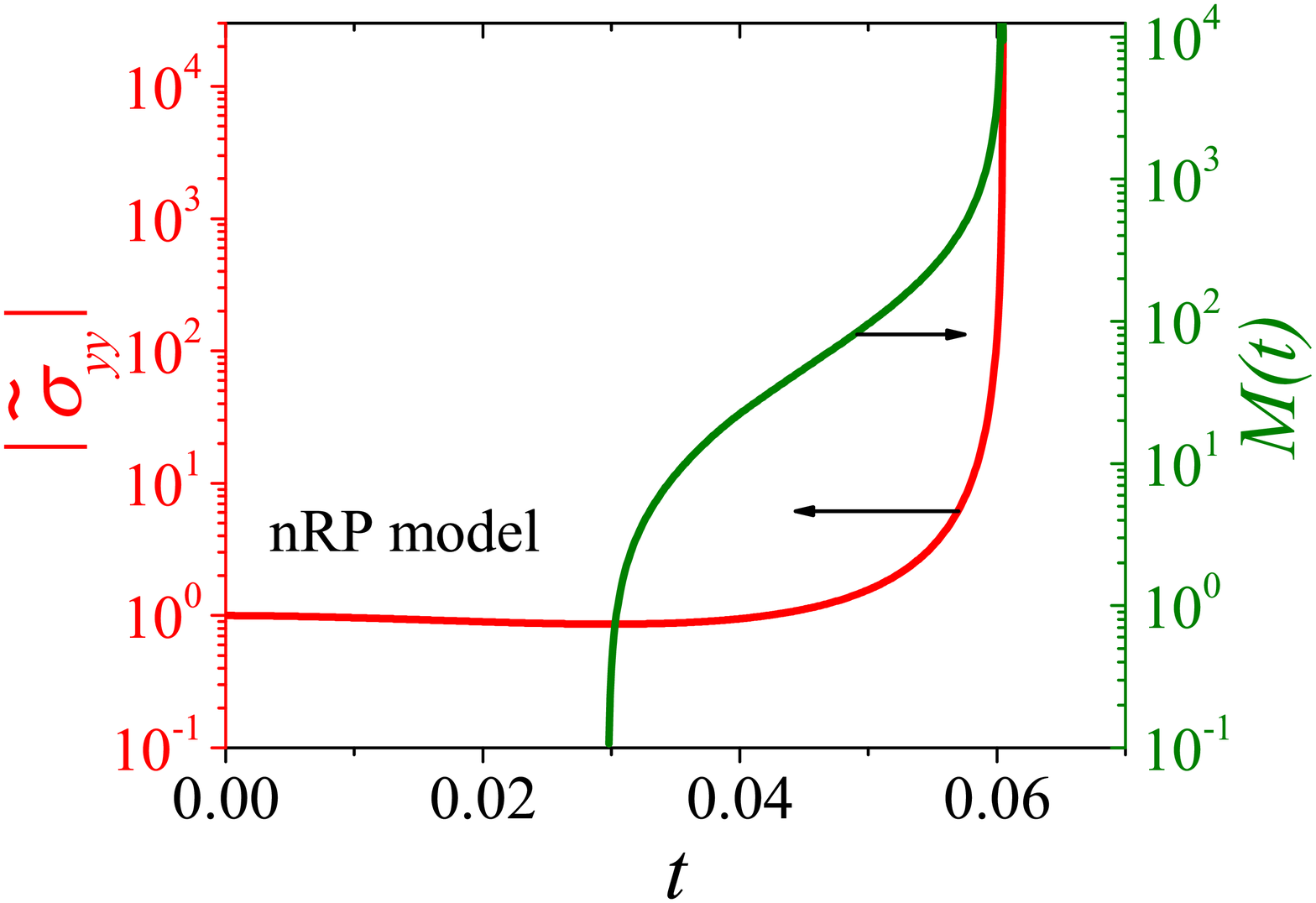}
    
    \label{fig:nrp_n_0_g_m}
    }
     \subfigure[]{
    \includegraphics[scale=0.28]{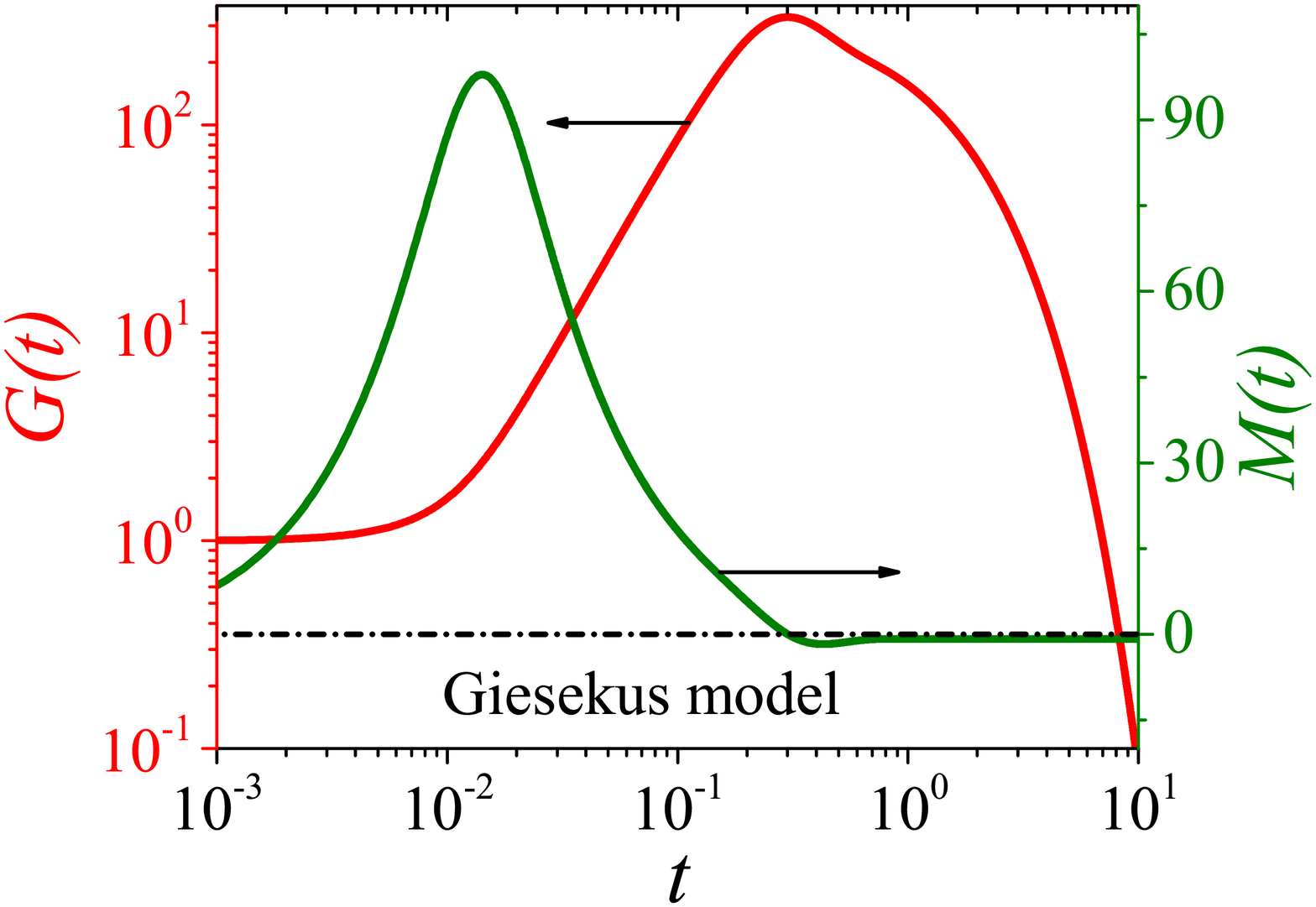}
    
    \label{fig:giesekus_n_0_g_m}
    }

    \caption{Evolution of $G(t)$ and $M(t)$ for (a) JS model using $Wi=7$ and $\eta_s=0$, (b) nRP model using $Wi=30$ and $\eta_s=0$, and (c) Giesekus model using $Wi=100$ and $\eta_s=0$. $G(t)$ for nRP model is represented by $|\overset{\sim}{\sigma}_{yy}|$ in the case of $\eta_s=0$. The perturbation growth coefficient diverges only for the case of nRP model.} 
    \label{n_0_results}
\end{figure}

The reason behind this contrasting effect of flatness of constitutive curve on JS and nRP model results is the value of $\eta_s$ and its effect on growth rate. Figure \ref{fig:eigenvalue_diverge} shows that the growth rate from the frozen-time analysis diverges as $1/\eta_s$ for $\eta_s \rightarrow 0$ for the viscoelastic models studied. Even the decay rate of the stable eigenvalue in the Giesekus model diverges in a similar manner. The divergence of decay rate is also observed in nRP and JS models for shear start-up in the non-flat regions of the constitutive curve. This implies that it is the limit of $\eta_s \rightarrow 0$, and not the flatness of the constitutive curve \textit{per se}, that imparts a large growth/decay rate (in the frozen-time analysis), in the viscoelastic models studied, and it is this feature that controls the growth rate (frozen-time analysis) and transient stability of shear start-up flow. Hence, the transient stability of a shear start-up flow on the basis of frozen-time analysis (which has its own limitations as discussed above) completely depends on $\eta_s$ for the studied viscoelastic models. 

An important feature regarding divergence of $\omega$ with decrease in $\eta_s$ is that it is not universal for the case of growth rate. Figure \ref{fig:eigenvalue_diverge_nRP} shows that for $\beta=1$, nRP model does not show divergence of $\omega$ and plateaus at lower value of $\eta_s$ at $Wi=10, 15, 17$, and $17.65$. This plateauing of $\omega$ is seen if $Wi<18$ for nRP model using $\beta=1$. Therefore, for sufficiently slow shear rates, there will not be any divergence of $\omega$. Similarly, the constitutive models that do not account for non-affine deformation, such as the Giesekus model, do not show divergence of transiently maximum growth rate with decrease in $\eta_s$. Interestingly, the scaling followed in this case before plateauing is $\omega\sim\eta_s^{-1/2}$ in contrast to the scaling of $\omega\sim\eta_s^{-1}$ which is followed if $Wi>18$.

%

Further, we discuss the effect of increasing the flatness of the constitutive curve on the linearized evolution of perturbation results for the JS, nRP and Giesekus models. The transient maximum value of $G(t)$ for the JS model shows a negligible change in response to varying $\eta_s$ as compared to the nRP model, while for the Giesekus model, it is almost constant. The peak value of $M(t)$ also shows a slight increase for the JS model as compared to the nRP model and it is almost constant for the Giesekus model on decreasing the value of $\eta_s$. This observation of the JS and the nRP models can be attributed to diverging transient eigenvalue with decrease in $\eta_s$, which is very large for the nRP model as compared to the JS model. Also, the order of magnitude of maximum value of $G(t)$ in Fig. \ref{fig:js_flat} (in which $t_p=0$) and Fig. \ref{js_tp} (in which $t_p>t_{steady-state}$) is similar. Therefore, the transient growth in the case of JS model cannot be a reliable indicator of transient shear banding in the linear regime. On the other hand, the transient growth in the case of nRP model in Fig. \ref{fig:nrp_flat} (in which $t_p=0$) is several orders of magnitude higher than for the case shown in Fig. \ref{nrp_tp} (in which $t_p>t_{steady-state}$). Under the creeping flow assumption, the significantly high value of actual growth rate and the diverging transient maximum value of $G(t)$ for the nRP model in Figs. \ref{lsa_m_nRP} and \ref{fig:nrp_flat_1e-5} shows a possibility of transient instability. (Whether a high transient growth can lead to transient instability or not, can only be determined using non-linear simulations). However, in the next section, we address a possible limitation of using the creeping flow assumption.

The linearized evolution results for the Giesekus and JS models also show that even if there is a stable transient eigenvalue, there can be growth and decay of perturbations as shown in Fig. \ref{fig:g_flat} as well as for the case of parameters of Fig. \ref{js3} (results not shown here; this is contrary to the transient shear banding criterion proposed in [\onlinecite{moorcroft14}], which is rooted in the frozen-time analysis, as this excludes transient, non-exponential, growth of perturbations for viscoelastic start-up shear flow when the eigenvalues are stable). The results for these two models reveal that if a perturbation is introduced to the time-dependent base state, then there will always be a growth of perturbations, even if the eigenvalue (within the frozen-time analysis) remains stable at all times, including at steady state. The transient nature of evolution of perturbation strongly depends on the base state, which varies from one constitutive model to the other. More importantly, the overshoot in the base-state shear stress (in the stress vs strain plots) does not trigger strong growth of perturbations for both the JS and Giesekus models, as shown by the peak values of $M(t)$. However, the transient maximum value of $G(t)$ is $O(10^2)$ and cannot be treated as a reliable precursor of substantial transient growth because the order of transient growth for the Giesekus model for the time-evolving base state is similar to the order of transient growth obtained if perturbation is seeded at time when base state has already achieved steady state as shown in Fig. \ref{giesekus_tp}. Therefore, if there is no significant effect of $\eta_s$ on the base state, then we shall find no effect of increasing the flatness of the constitutive curve on the evolution of $G(t)$ and $M(t)$. Finally, this discussion suggests that a quantitative measure of transient shear banding can be decided on the basis of the ratio of maximum value of $G(t)$ when $t_p>t_{steady-state}$ to the maximum value of $G(t)$ when $t_p=0$ or $t_p \ll t_{steady-state}$.

Interestingly, instead of studying $\eta_s \rightarrow 0$, the results of $G(t)$ and $M(t)$ can also be obtained directly using $\eta_s=0$ for the JS, nRP and Giesekus models. Figure \ref{n_0_results} shows that $G(t)$ diverges for the case of nRP model using $\eta_s=0$, while the results for the JS and Giesekus models do not show any divergence. The finite-time blow up in case of nRP model is consistent with the divergence of $\omega$ with decrease in $\eta_s$. However, the results using $\eta_s=0$ are strictly relevant only from a theoretical standpoint for the following reason. The value of $\eta_s$ cannot be strictly equal to zero for polymeric melts and concentrated polymer solutions, since a small nonzero $\eta_s$ is usually considered as a proxy for faster relaxing degrees of freedom [\onlinecite{moorcroft14,adams2011transient,rallison1997dissipative}]. The three models used in this study are single-mode models and the faster relaxing modes can be accounted for by the $\eta_s$ term, which does not merely represent the presence of solvent (as shown in the Ref. [\onlinecite{rallison1997dissipative}] for extensional flow).


It must be noted that, as mentioned in the beginning of Sec. \ref{section_model}, there are inherent limitations to a linearized study, and nonlinear simulations are the most reliable tools to study transient shear banding. There are also several issues with regard to providing appropriate magnitude of initial conditions for nonlinear simulations, and these will be addressed in a subsequent communication.

\subsection{Effect of inertia} \label{inertia_effect}

We next study the effect of fluid inertia on the linearized evolution for all the three models considered here. Strictly speaking, with the inclusion of fluid inertia, one must consider the base state velocity profiles also in the presence of inertia, which evolve over the timescale of $H^2/\nu$ to reach steady state $\left(\nu=\frac{\eta_s+\eta_p}{\rho}\right)$ [\onlinecite{denn1971elastic}]. Such an analysis would be very involved, and we defer this for a future study. However, when there is a time-scale separation between the time required for the velocity profile to reach linearity, and for the stresses to reach their steady state values, then it is self-consistent to neglect inertia in the base-state, but include it in the linearized evolution. The ratio of the two timescales is the elasticity number $E = \lambda /(H^2/\nu) = Wi/Re$. When $E \gg 1$, then the assumption of neglecting inertia in the base-state is justifiable. In the following analysis, we carry out the frozen-time eigenvalue analysis and the linearized evolution of perturbations in the presence of fluid inertia, especially in the limit of $\eta_s \rightarrow 0$. In this limit, as demonstrated earlier in this paper, the time rate of change of velocity diverges, and hence inertial terms in the Cauchy momentum balance start becoming dominant. In other words, the presence of inertia `regularizes' the solution of linearized evolution of perturbation during shear start-up flow because in absence of inertia, Eq. \ref{cauchy_momentum_pert_inertia} reduces to an algebraic equation.  

\begin{figure}
    \subfigure[]{
    \includegraphics[scale=0.28]{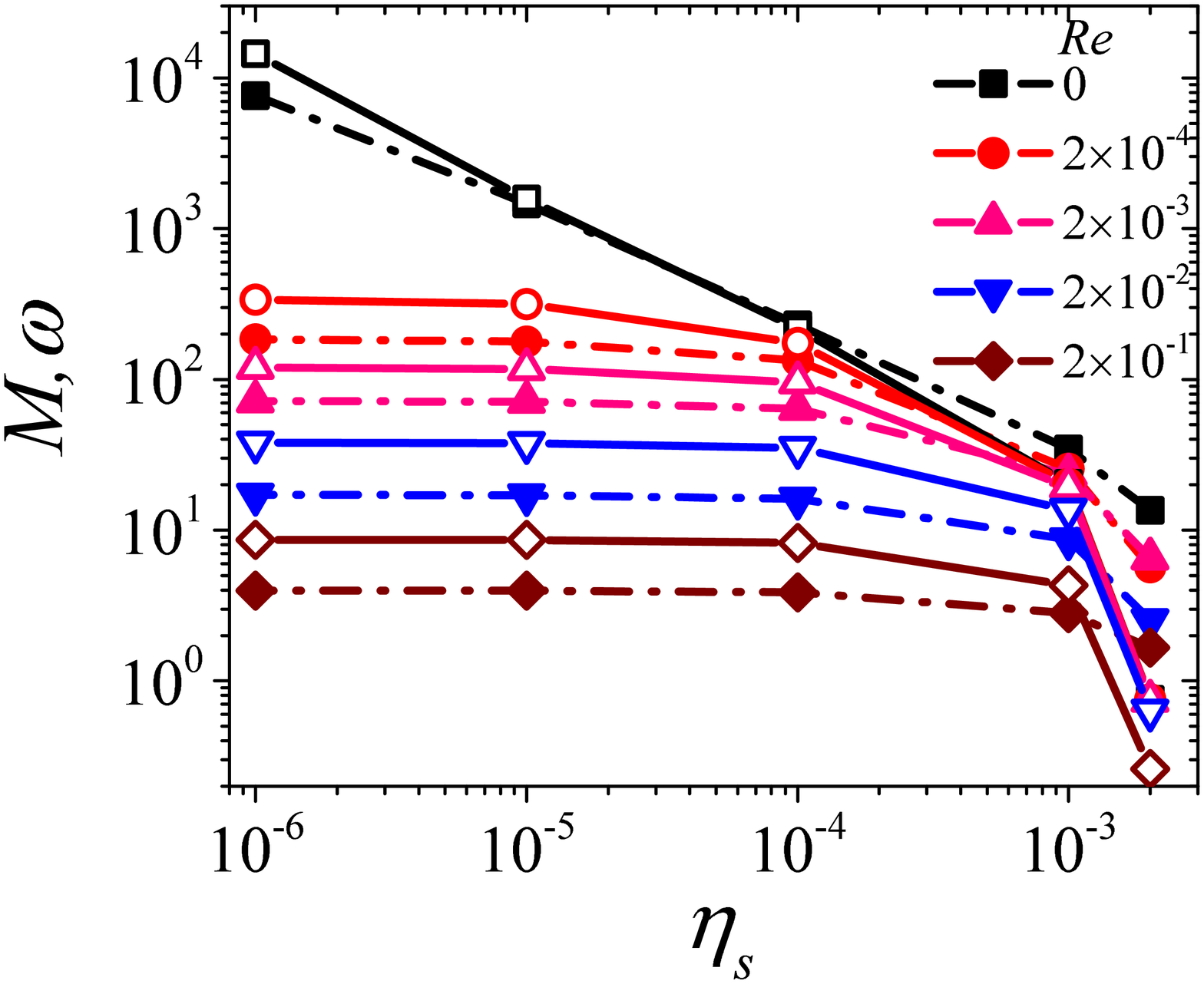}
   
    \label{fig:nrp_eigen_inertia}
    }
    \subfigure[]{\includegraphics[scale=0.28]{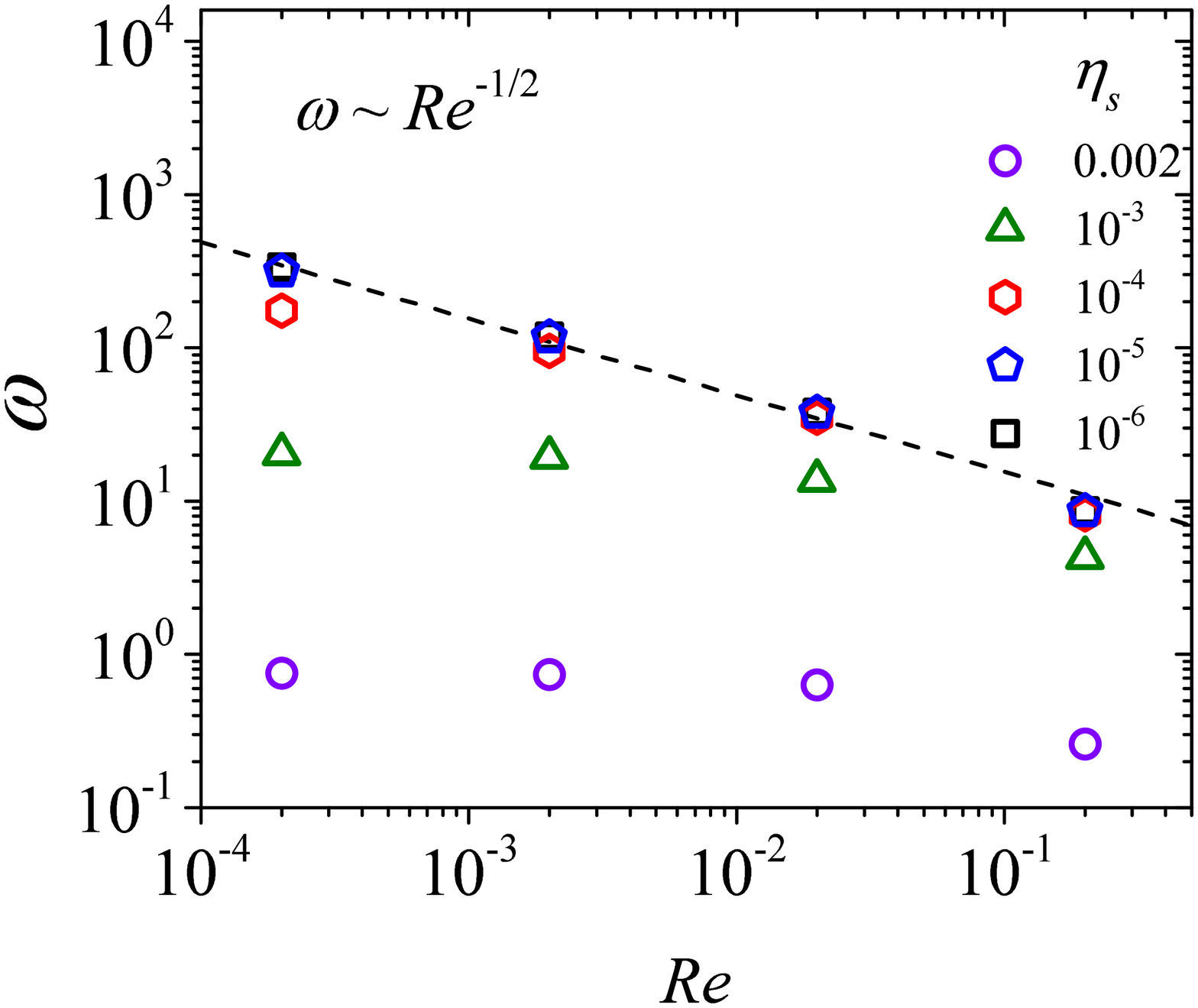}
    
     \label{fig:omega_Re}
     }
     \caption{(a) Effect of $\eta_s$ and $Re$ on transiently maximum value of eigenvalue $\omega$ (open symbols) and growth rate,~$M(t)$ (closed symbols) at $Wi=30$ and $t=0.07$ for nRP model with $\beta=1$. The solid and dashed lines are guide to eyes. The inclusion of inertia saturates the divergence of transient maximum value of $\omega$, and $M(t)$ with decrease in $\eta_s$. (b) Variation of $\omega$ is plotted as a function of $Re$ for different values of $\eta_s$. The dashed line shows the agreement of asymptotic relation $\omega \sim Re^{-1/2}$ with the numerical results in the limit of small $\eta_s$.}
\end{figure}

We find that results from the frozen-time analysis for the JS model are identical in the presence or absence of inertia. However, the nRP model results shown in Fig.~\ref{fig:nrp_eigen_inertia} depict the saturation of transiently maximum value of eigenvalue ($\omega$) and actual growth rate $(M)$ on decreasing $\eta_s$ for $Re=2 \times 10^{-4}$ as compared to divergence of eigenvalue for $Re=0$. On increasing $Re$ further, $\omega$ saturates to relatively lower value. The transiently maximum value of $\omega$ is appreciably high even in presence of inertia, which may or may not lead to the transient shear banding. This can be ascertained only by nonlinear simulations, and will be discussed in a future communication. Interestingly, solving Eq. \ref{cauchy_momentum_pert_inertia} in the limit of $\eta_s\rightarrow0$, the following asymptotic result can be obtained:
 \begin{equation}
     \omega\sim Re^{-1/2} \mbox{ if } \eta_s\rightarrow0
 \end{equation}
Therefore, at very low values of $\eta_s$, the value of $\omega$ will be of the order of $Re^{-1/2}$. Interestingly, it also corroborates with numerical results as shown in Fig. \ref{fig:omega_Re}.
 
 There is no transiently positive eigenvalue for the case of Giesekus model, therefore, the effect of inertia is subdominant. Fluid inertia has no effect on linearized evolution of perturbations, growth coefficient and actual growth rate for JS and Giesekus models because of the low growth rates obtained for the JS model and only stable eigenvalues for Giesekus model. The effect of inertia on linearized evolution of perturbations growth coefficient and actual growth rate for nRP model is shown in Fig. \ref{fig:nrp_flat}.

The transient maximum value of $G(t)$ in presence of inertia for nRP model decreases from $O(10^{30})$ (diverging) to $O(10^3)$ for the case of $\eta_s=10^{-5}$ as shown in Fig. \ref{fig:nrp_flat_1e-5}, and for the case of $\eta_s=10^{-3}$,~ $G(t)$ decreases from $O(10)$ to $O(1)$ as shown in Fig. \ref{fig:nrp_flat_1e-3}.

It is clear from the above discussion that inertia has an appreciable effect on the transient eigenvalue and $M(t)$ only for the nRP model. The growth rate from the frozen-time analysis and actual growth rate from the linearized evolution of perturbation growth coefficient are in good agreement in presence of inertia and this is similar to the results under creeping flow approximation discussed above for the nRP model. The reason behind this observation is the transiently high value of $\omega$ (which is $\sim O(100-1000)$) for the nRP model as compared to JS model for which $\omega \sim O(0.1-1)$. 
If the value of $\eta_s$ is of the order of $10^{-5}-10^{-4}$, then the magnitude of $\omega$ diverges in that limit. Thus, although the inertial effects in concentrated polymer solutions (owing to their large viscosity) are nominally negligible, they cannot be neglected when the growth rates themselves diverge for $\eta_s \rightarrow 0$. Therefore, neglect of inertia yields unphysical results both in frozen-time analysis as well as in linearized evolution of perturbations and inclusion of inertial terms even with $Re = 2 \times 10^{-4}$ has a dramatic impact on the eigenvalue and $G(t)$, resulting in modest transient growth in terms of magnitude of growth and decay of perturbations. Additionally, all the results of nRP model are in the absence of chain-stretching as per the model constitutive equations. In the absence of inertia and small value of $\eta_s$, addition of chain-stretching may have a different dynamics altogether and play an important role. The stretching Rolie-Poly model, although, has been studied thoroughly by Adams and Olmsted [\onlinecite{adams2009nonmonotonic,adams2011transient}] in the context of transient shear banding. The effect of decreasing $\eta_s$, in presence of chain-stretching and inertia may provide a new insight.

However, this is an approximate analysis as it considers that the base state reaches a homogeneous velocity profile in time, and the momentum diffusion time is considered to be at least an order of magnitude lesser than the effective relaxation time of the solution. A more involved analysis with base state also incorporating inertia while evolving to the steady state will be discussed in future work. While the present study focuses on a small nonzero inertia as a way to mitigate unlimited growth, it must be mentioned that other physical processes (not considered in this work) such as faster relaxation processes, 3D flows etc., can also potentially play a role in mitigating unlimited growth.

As discussed in the Introduction, there are reports which appear to show agreement [\onlinecite{adams2011transient,mohagheghi2016elucidating,cao2012shear,adams2009nonmonotonic,zhou2008modeling,zhou2012multiple,zhou2014wormlike,tapadia2006direct,ravindranath2008banding,wang2011homogeneous,wang2014letter,boukany2008use,boukany2009exploring,boukany2009shear,moorcroft14}] with the transient shear banding criteria  [\onlinecite{moorcroft14}], and studies which show disagreement [\onlinecite{li2013flow,ianniruberto2017shear,benzi2021continuum}] with the universal nature of the same.  In the present effort, we studied three widely used models, viz., JS, nRP and Giesekus, and critically examined the agreement between our predictions and the criterion for onset of transient banding during shear start-up. We find that while the nRP model predictions agree well with the criterion [\onlinecite{moorcroft14}] in some parameter regimes, the predictions from the JS and Giesekus models do not agree with the criterion for the onset of transient shear banding. This suggests that the criterion for onset of transient banding during shear start-up is not fluid universal. It will be useful to explore other constitutive models to obtain a more comprehensive understanding of transient banding.

\section{Conclusions} \label{section_conclusions}

We have carried out linear stability analyses to understand the transient dynamics during shear start-up flow of a viscoelastic fluid using Johnson-Segalman, non-stretching Rolie-Poly and Giesekus models. 
We considered shear rates in monotonic regions of the constitutive curves with a focus on the onset of transient shear banding during shear start-up flow. We revisit the criterion proposed [\onlinecite{moorcroft14}] for the onset of transient shear banding in the context of shear start-up flow, which was based on an analysis rooted in the  frozen-time eigenvalue analysis. We argue that, even within this framework, the criterion must be viewed as a necessary, but not sufficient, condition for the onset of transient banding. This implies that not all constitutive models that obey the criterion will exhibit a transient instability. We further show that, within the context of the frozen-time analysis, the eigenvalue is not a reliable marker to ascertain transient stability of the system because of the lack of a general correlation between the nature of the eigenvalue and the growth and decay of perturbations, the latter being obtained numerically using the fundamental matrix approach. The presence of Hopf bifurcation during the transient evolution in the Johnson-Segalman model further reinforces the restrictive nature of the transient shear banding criterion proposed in the literature [\onlinecite{moorcroft14}], which explicitly ignored this possibility. Moreover, results from our linear stability analysis also suggest the lack of any universal trend between shear stress overshoot and a positive eigenvalue, and we attribute this to the criteria derived earlier being necessary rather than sufficient conditions for transient instability. The disagreement between the criterion for transient shear banding [\onlinecite{moorcroft14}] and the results obtained using the JS and Giesekus models could also be attributed to the fact that the criterion was derived for models with only two dynamic viscoelastic variables; whereas, the JS and Giesekus models have three dynamic viscoelastic variables.

We also discussed in detail the subtleties involved in taking the limit of $\eta_s \ll 1$, a limit that was frequently considered in earlier efforts on transient shear banding [\onlinecite{adams2011transient,moorcroft14}].
Firstly, the growth rate of the unstable eigenvalue diverges for $\eta_s \ll 1$, but this limit also results in increasing the flatness of the constitutive curve. Earlier efforts have attributed this flatness of the constitutive curve to large transient growth. However, we demonstrate that it is the decrease in $\eta_s$, and not the flatness of the constitutive curve per se, that resulted in a divergent growth rate in the frozen-time analysis. We illustrate this via a counter example using the JS model where $\eta_s$ could not be arbitrarily decreased because the constitutive curve becomes non-monotonic below $\eta_s = 1/9$. For this model, there is no substantial increase in the transient growth even as the flatness is increased within the parametric regimes where the constitutive curve is monotonic. In contrast, for low $\eta_s$ (which yields a flatter constitutive curve),  while the Giesekus model yields a negative transient eigenvalue, the linearized evolution shows a growth and eventual decay of perturbations, as determined by the growth coefficient $G(t)$. We also illustrated the importance of retaining fluid inertial effects in the limit of $\eta_s \ll 1$, because the unstable eigenvalue diverges as $1/\eta_s$ in that limit. Inclusion of inertia was accomplished in an approximate manner, wherein it was considered only in the linearized evolution of perturbations, while the base-state was calculated within the creeping-flow approximation. Inclusion of small, but finite inertia decreases the maximum value of perturbation amplification by several orders of magnitude for the nRP model and does not affect JS and Giesekus model results because of low magnitude of growth rate (JS) or decaying nature (Giesekus) of $\omega$.

In conclusion, our results provide a more comprehensive and definitive support to the the notion [\onlinecite{fielding2016triggers}] that the criterion proposed in the literature [\onlinecite{moorcroft14}] for the onset of transient banding during shear start-up may not be fluid-universal. Unlike steady-state banding, for which there is the well-known `negative-slope' criterion in the stress-strain rate curve [\onlinecite{yerushalmi1970stability}], there appears to be no analogous, overarching, yet simple, criterion for the formation of transient bands during shear start-up of fluids with monotonic constitutive curves.
Thus, based on the JS, nRP and Giesekus models considered in the present study,  it is reasonable to conclude that, for monotonic constitutive curves, the presence of a stress overshoot in the stress-strain curve after shear start-up does not necessarily result in the onset of an elastic instability leading to onset of pronounced transient banding if (i) the value of $\eta_s$ is not so low such that $\omega\sim\eta_s^{-1}$, and (ii) fluid inertial effects are included, especially for $\eta_s \ll 1$.
More work, with other constitutive models, is required to ascertain the universal nature of this proposition.  
There are other possible physical mechanisms that could potentially mitigate the finite-time blow-up of the nRP model, such as the inclusion of a finite stretch relaxation time or 2D flow effects, which can be interesting avenues for future research.
 
\section*{Acknowledgment}
We acknowledge financial support from the Science and Engineering Research Board, Government of India. We also thank Morton Denn, Gareth McKinley, Peter Olmsted, Thibaut Divoux and Xue-Feng Yuan for their insightful comments and critical reading of the manuscript. We are particularly thankful to Joseph D. Peterson for specifically suggesting us to present results related to 
Fig. \ref{fig:eigenvalue_diverge_nRP},  and for other insightful comments on the manuscript. 




\bibliography{mybibfile}

\end{document}